\documentclass[10pt]{article}
\usepackage[OE]{express}
\usepackage{subfigure}
\usepackage{color}

\def\um{$\mu$m}

\begin{document}
\title{Photonic ring resonator filters for astronomical OH suppression}


\author{S.~C.~Ellis,\authormark{1,2,*} S.~Kuhlmann,\authormark{3} K.~Kuehn,\authormark{1} H.~Spinka,\authormark{3} D.~Underwood,\authormark{3} R.~R.~Gupta,\authormark{3} 
L.~Ocola\authormark{3}, P.~Liu,\authormark{4} G.~Wei,\authormark{4} N.~P.~Stern,\authormark{4}
J.~Bland-Hawthorn,\authormark{2}  and P.~Tuthill\authormark{2} }

\address{
\authormark{1}Australian Astronomical Observatory, 105 Delhi Rd., North Ryde, NSW 2113, Australia\\
\authormark{2}Sydney Institute for Astronomy, School of Physics, University of Sydney, NSW 2006, Australia\\
\authormark{3}Argonne National Laboratory, 9700 S.~Cass Avenue, Argonne, IL 60439, USA\\
\authormark{4}Northwestern University, 2145 Sheridan Road, Evanston, IL 60208, USA
}

\email{\authormark{*}sellis@aao.gov.au} 



\begin{abstract}
Ring resonators provide a means of filtering specific wavelengths from a waveguide, and optionally dropping the filtered wavelengths into a second waveguide.  Both of these features are potentially useful for astronomical instruments.

In this paper we focus on their use as notch filters to remove the signal from atmospheric OH emission lines from astronomical spectra, however we also briefly discuss their use as frequency combs for wavelength calibration and as drop filters for Doppler planet searches.

We derive the design requirements for ring resonators for OH suppression from theory and finite difference time domain simulations.   We find that rings with small radii ($<10$~$\mu$m) are required to provide an adequate free spectral range, leading to high index contrast materials such as Si and Si$_{3}$N$_{4}$.  Critically coupled rings with high self-coupling coefficients should provide the necessary $Q$ factors, suppression depth, and throughput for efficient OH suppression.

We report on our progress in fabricating both Si and Si$_{3}$N$_{4}$ rings for OH suppression, and give results from preliminary laboratory tests.  Our early devices show good control over the free spectral range and wavelength separation of multi-ring devices.  The self-coupling coefficients are high ($>0.9$), but further optimisation is required to achieve higher $Q$ and deeper notches, with current devices having $Q\approx 4000$ and $\approx 10$~dB suppression.  
The overall prospects for the use of ring resonators in astronomical instruments is promising, provided efficient fibre-chip coupling can be achieved.
\end{abstract}





\bibliographystyle{osajnl}
\bibliography{ps}

\section{Introduction}
\label{sec:introoh}

Observations at near-infrared (NIR) wavelengths (0.9 -- 2.5 \um) are crucial for many areas of astronomy.  For example, the lowest mass stars and highest mass planets emit most of their light at near-infrared wavelengths, and NIR spectroscopy is essential for classifying such objects.   Dusty regions within our own and other galaxies are highly opaque to visible wavelengths, but transparent at long wavelengths due to the $\lambda^{-4}$ dependence of Rayleigh scattering.  Thus, studying the inner regions of the Milky Way, or star-forming regions within other galaxies, often requires NIR spectroscopy.      Deep NIR spectroscopy is also necessary to study the high redshift Universe, when the diagnostic visible features are redshifted into the NIR.
Measuring star-formation rates during the epoch of peak star-formation, measuring Lyman-$\alpha$ emission during the epoch of reionisation, and identifying high redshift supernov\ae\ would all benefit significantly from NIR spectroscopy.

\begin{figure}
\centering \includegraphics[scale=0.5]{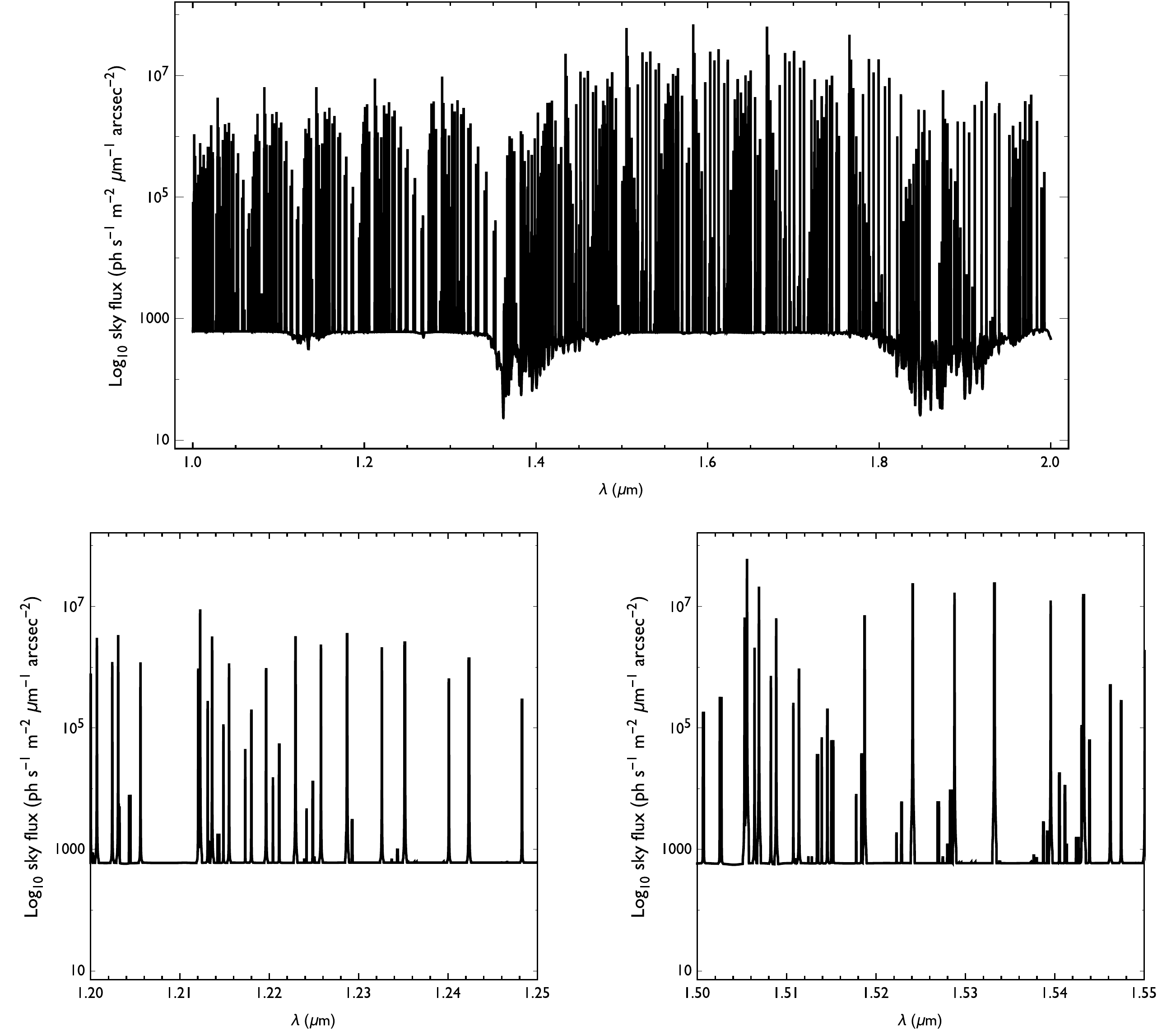}
\caption{A model spectrum of the near-infrared night sky.  The background is dominated by OH emission lines (top plot).  The OH lines are intrinsically very narrow, with dark sky in between.}
\label{fig:nirsky}
\end{figure}

Unfortunately, deep observations at near infrared wavelengths are currently very challenging due to the bright atmospheric background.  The surface brightness of the night sky at a good site is thousands of times brighter in the NIR than in the visible.
Between $0.9$ --  1.8~\um\ almost all of this background results from the de-excitation of atmospheric OH molecules\cite{mei50,duf51} at an altitude of $\approx 90$ km giving rise to an extremely bright emission line spectrum, shown in Figure~\ref{fig:nirsky} (see Ellis \& Bland-Hawthorn.\cite{ell08} and references therein for a review of the NIR background). Not only is this OH line spectrum extremely bright, it is also variable on a time scale of minutes. Subtracting this background from astronomical observations is very inaccurate due to the large Poissonian and systematic noise\cite{dav07}. Solving the difficulty of the NIR night sky background is a long standing problem in astronomy.

Several solutions to this problem have been proposed.  One obvious, but expensive, solution is to launch a telescope into orbit above the atmosphere.  Indeed, this is one of the primary motivations for the James Webb Space Telescope (JWST)\cite{gar06}, which is planned for launch in 2018 at a cost of $\approx$~US\$ 8.8 billion.  While NIR space telescopes completely avoid the problem of atmospheric emission, ground-based solutions are also desirable: space telescopes have a finite lifetime and are expensive to replace; ground-based facilities can take advantage of developments in technology and scientific understanding and allow for the optimisation of instruments for specific requirements as the need arises.  Furthermore, in the case of JWST the field-of-view of the  NIRSPEC instrument is $3.6 \times 3.4$~arcmin\cite{bir16}, significantly limiting its ability to perform large surveys.  Additionally, space-based observatories are often significantly more over-subscribed than ground-based facilities, leading to significant restrictions in the proportion of researchers able to take advantage of such resources.

Ground-based solutions exploit the fact that the OH lines are intrinsically very narrow (FWHM $\approx 0.3$~pm), with a fixed wavelength, and the sky between the lines is very dark, as shown in the bottom panels of Figure~\ref{fig:nirsky}.  If the lines can be filtered at high resolution and high suppression, then the remaining background spectrum will be orders of magnitude darker.

It is not possible  simply to observe at high resolution to see between the lines, since the OH lines will be smeared by the spectrograph scattering point spread function, contaminating the inter-line region.  To realise efficient OH suppression this scattering problem must be avoided; the OH light must be removed prior to dispersion.  Furthermore, it is often desirous to observe at lower spectral resolution to obtain better signal-to-noise and larger wavelength coverage.


Several ground-based solutions have been proposed.  These include high-dispersion masking\cite{mai93b,iwa94,mai00,iwa01,mot02}, ultra-narrow band filters\cite{hor04}, Rugate filters\cite{off98} and holographic filters\cite{blai04}.  All of these have met with limited success, either due to the spectrograph scattering problem, difficulties in fabrication or difficulties in implementation (see Ellis \& Bland-Hawthorn.\cite{ell08} for a full discussion).



A more recent solution uses fibre Bragg gratings (FBGs) \cite{bland04,bland11b}, wherein up to 150 individual notches may be written into a single fibre\cite{bland08}, which are perfectly matched in wavelength to the OH lines, have up to 40~dB of suppression, and a width of $\approx 1.5$~\AA.  Such devices have been proven on-sky with the GNOSIS prototype instrument\cite{ell12a,tri13a,tri13b}.   The chief limitation of FBGs is that they must be written into individual single mode fibres, and therefore require a photonic lantern\cite{leo05,noo09} to couple efficiently to a spectrograph.  Even so, FBGs are currently only suitable for single object spectroscopy, since the cost of scaling GNOSIS-type instruments to larger field of view is currently prohibitive.  More efficient methods of manufacture are currently being investigated, for example multicore FBGs\cite{bir12,lin14} would simplify the production and integration of large numbers of fibres.

\begin{figure}
\centering \includegraphics[scale=0.4]{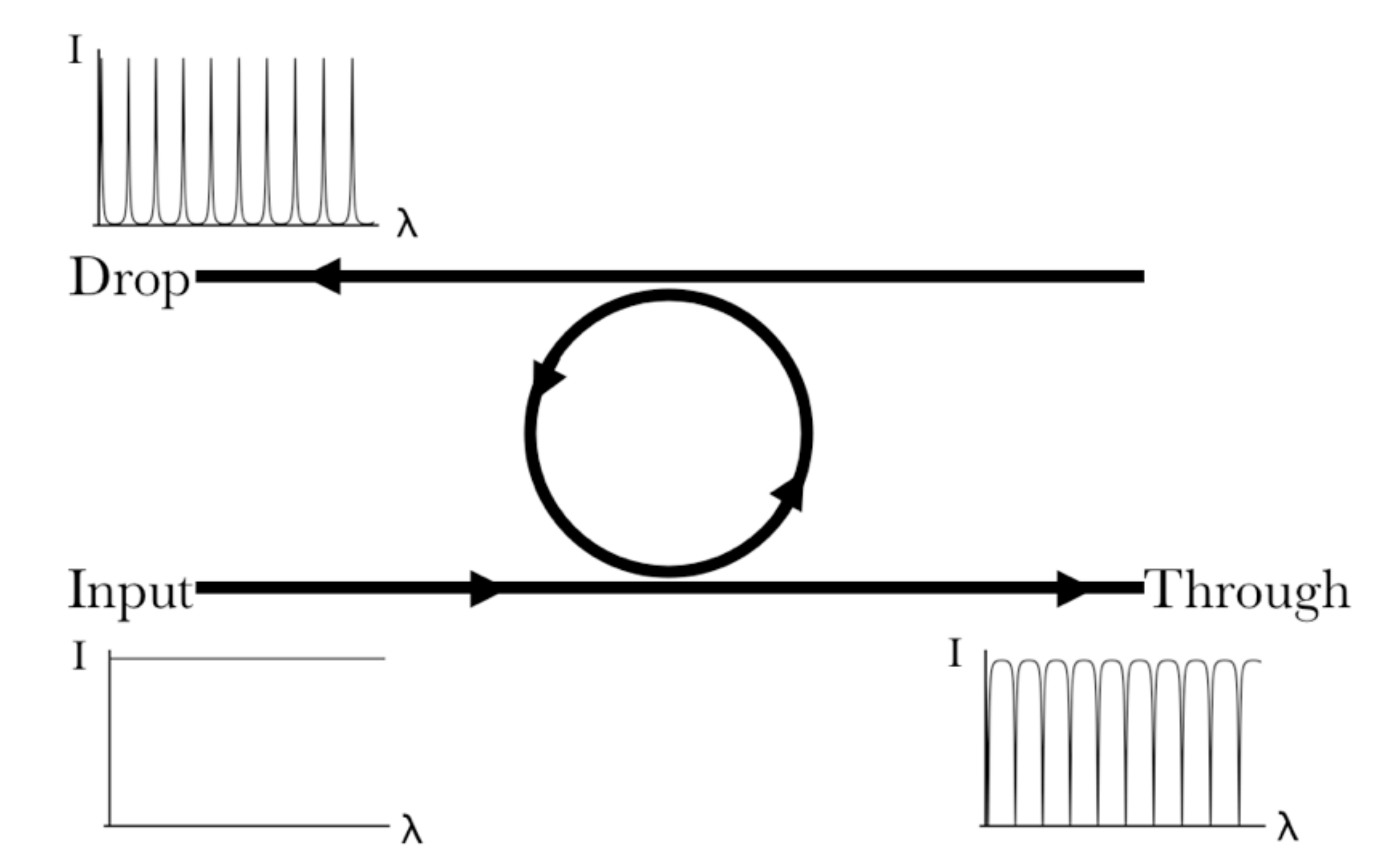}
\centering \includegraphics[scale=0.5]{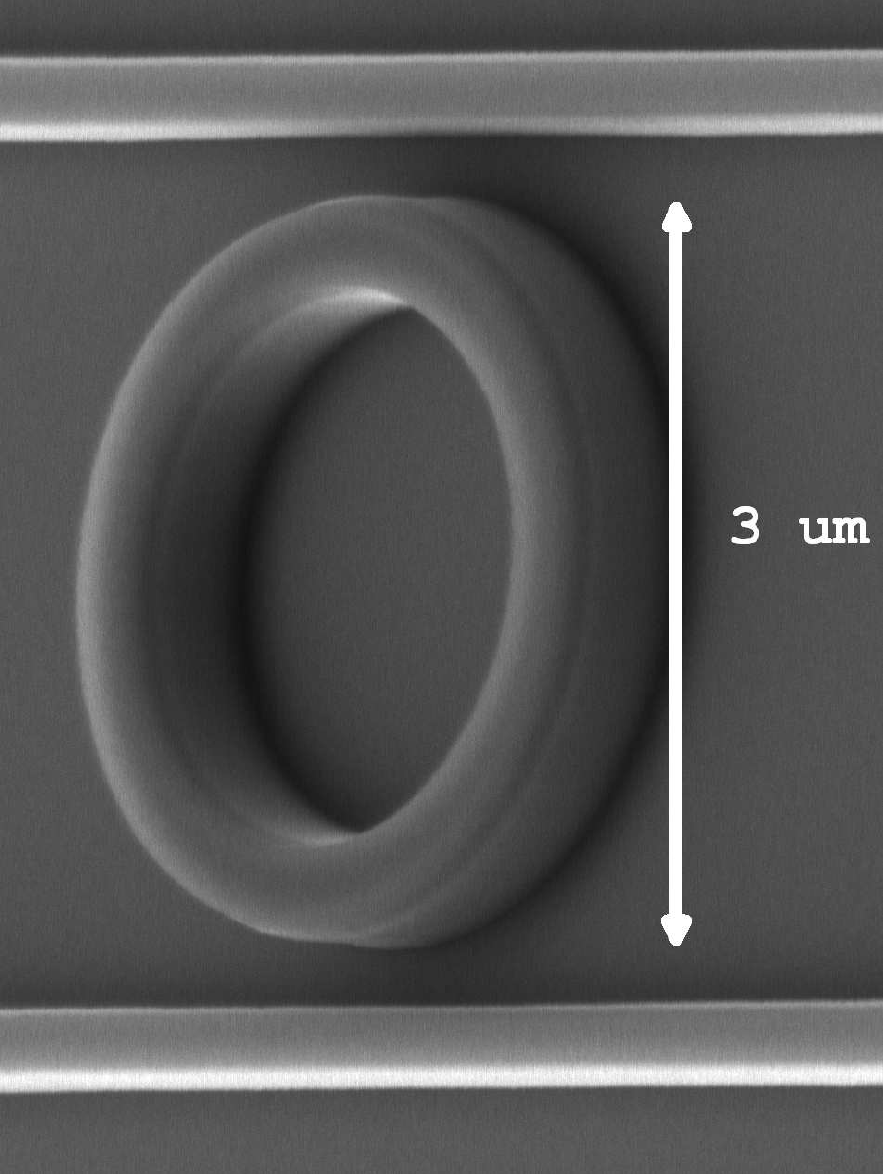}
\caption{Top panel: schematic diagram of a simple ring resonator showing the input, through and drop ports and a sketch of the spectrum at each port.  Bottom panel:  SEM photograph of a silicon-based ring resonator with a through port and drop port.}
\label{fig:schematic}
\end{figure}

In this paper we examine another class of notch filter, viz.\ ring resonators.  Ring resonators consist of one or more waveguides coupled to a looped waveguide -- see Figure~\ref{fig:schematic}.  Light from the input waveguide evanescently couples into the ring, whereupon it constructively and destructively interferes with itself until only light at the resonant wavelengths of the cavity remains.  The condition for resonance is therefore
\begin{equation}
\label{eqn:resonance}
m \lambda = n_{\rm e} L,
\end{equation}
where $m$ is an integer, $\lambda$ is the wavelength, $n_{\rm e}$ is the effective index
and $L$ is the circumference of the ring.  The resonant light couples back into the input waveguide and destructively interferes with the input light.  
Thus, a series of ring resonators, each tuned to the wavelength of a different OH night sky line, could provide a means of OH suppression\cite{ell12c}.

Ring resonators have been developed for applications in telecommunications, industry, and photonics research as filters, add/drop multiplexers, delay lines, modulators, sensors, laser generation, tuneable dispersion compensators, all-optical wavelength converters, frequency combs, and tuneable cross-connects\cite{ven06,rab07}.  The motivation to explore the use of ring resonators for OH suppression is primarily due to their method of manufacture. Ring resonators can be lithographically printed in a photonic integrated circuit (PIC), and are therefore very versatile and repeatable.  They are extremely modular, potentially offering a much more scalable means of production than is possible using FBGs.  Ring resonators are inherently small devices ($1$ -- 100$\mu$m), providing an excellent means of miniaturising parts of instruments. Lithography enables many rings to be coupled to the same waveguide, therefore the insertion losses are limited to a single coupling loss at input and output, even for an arbitrarily high number of rings. Finally, since they are printed in a PIC, they may be easily combined with other photonic components, e.g.\ array-waveguide  gratings\cite{wat95,wat97,cve09,bland10} to form fully photonic systems.

This manuscript will describe the potential uses and challenges for ring resonators and report on early progress in laboratory tests of prototype devices.
In section~\ref{sec:ohsupp} we give an overview of the applicability of ring resonators for OH suppression.  Following this we develop the requirements for ring resonators in section~\ref{sec:ohreq}, based on their theoretical properties.  Then in section~\ref{sec:prac} we discuss the practical aspects of applying  ring resonators to OH suppression, including a discussion of our fabrication of Si and Si$_{3}$N$_{4}$ devices and our laboratory tests of these devices.  These results are discussed with reference to the requirements laid out in section~\ref{sec:ohreq}.   In section~\ref{sec:discuss} we summarise the feasibility of using ring resonators in astronomical instruments, and discuss the  future  development and testing necessary.   

Finally, we note that the interest in ring resonators for astronomy is not limited to OH suppression.  The light from the ring can be dropped into a second waveguide providing a comb of frequencies at the resonant wavelengths of the ring.  This dropped signal can be used to generate a frequency comb for wavelength calibration. The drop port can also be used to siphon particular wavelengths of interest, and to add several such signals together.  This could be useful for time domain astronomy, in particular for Doppler planet searches. We  discuss the potential of these other applications in brief in appendix~\ref{sec:other}.

\section{Ring resonators for OH suppression}
\label{sec:ohsupp}

The theoretical properties of ring resonators are well understood\cite{rab07}.  Using these properties we now quantify the astronomical requirements and the consequent design requirements on the use of ring resonators  for OH suppression.
In this section we give a brief depiction of how ring resonators can be incorporated into an near-infrared spectrograph to provide OH suppression (\S~\ref{sec:ohsuppintro}).  We then examine the astronomical requirements and the consequent technical requirements in section~\ref{sec:ohreq}.  

\subsection{An outline of OH suppression with ring resonators}
\label{sec:ohsuppintro}

\begin{figure}
\centering
\includegraphics[scale=0.7]{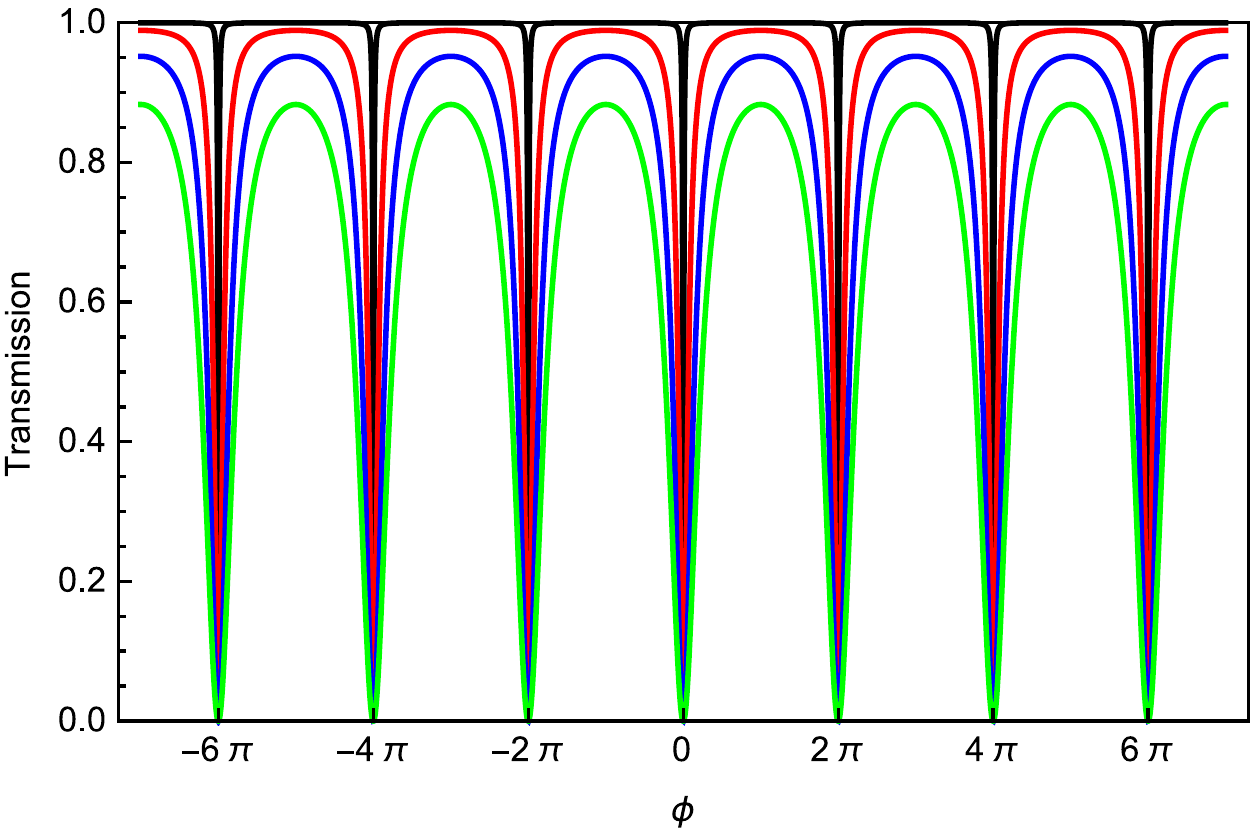}
\caption{The transmission for a simple ring resonator with no drop port as a function of phase with $\alpha=t=0.99$ (black), $\alpha=t=0.9$ (red), $\alpha=t=0.8$ (blue), and $\alpha=t=0.7$ (green).}
\label{fig:simplenotchthrough} 
\end{figure}

The general problem of OH suppression was introduced in section~\ref{sec:introoh}.  
We require a filter with a deep and narrow notch at the wavelength of each bright OH line.  (Actually the OH lines are doublets, most of which are very closely spaced such that a pair of lines can be suppressed with a single notch).  
Figure~\ref{fig:simplenotchthrough} shows  an example of the theoretical transmission of  a simple ring resonator with no drop port, which is critically coupled such that the throughput of one passage around the ring, $\alpha$, is equal to the self-coupling along the input waveguide  $t$.  The total phase change for a single passage around the ring may be expressed in terms of wavelength as, $\theta  = 2 \pi n_{\rm e} L/\lambda$.

Thus, light from the telescope can be fed into a PIC within which ring resonators provide this filtering function.  The filtered light from the through port of the ring resonators can then be fed into a spectrograph.

The OH lines are not periodic with frequency, whereas the multiple notches of a single ring resonator are.  Therefore each wavelength to be suppressed must be suppressed by an individual ring; if there are $N$ lines to be suppressed, there must be $N$ rings.  In the most optimistic case, it may be possible to suppress two lines with an individual ring, if the FSR can be matched to the spacing between the lines without compromising the $Q$ factor and suppression factor.

\begin{figure}
\centering \includegraphics[scale=0.5]{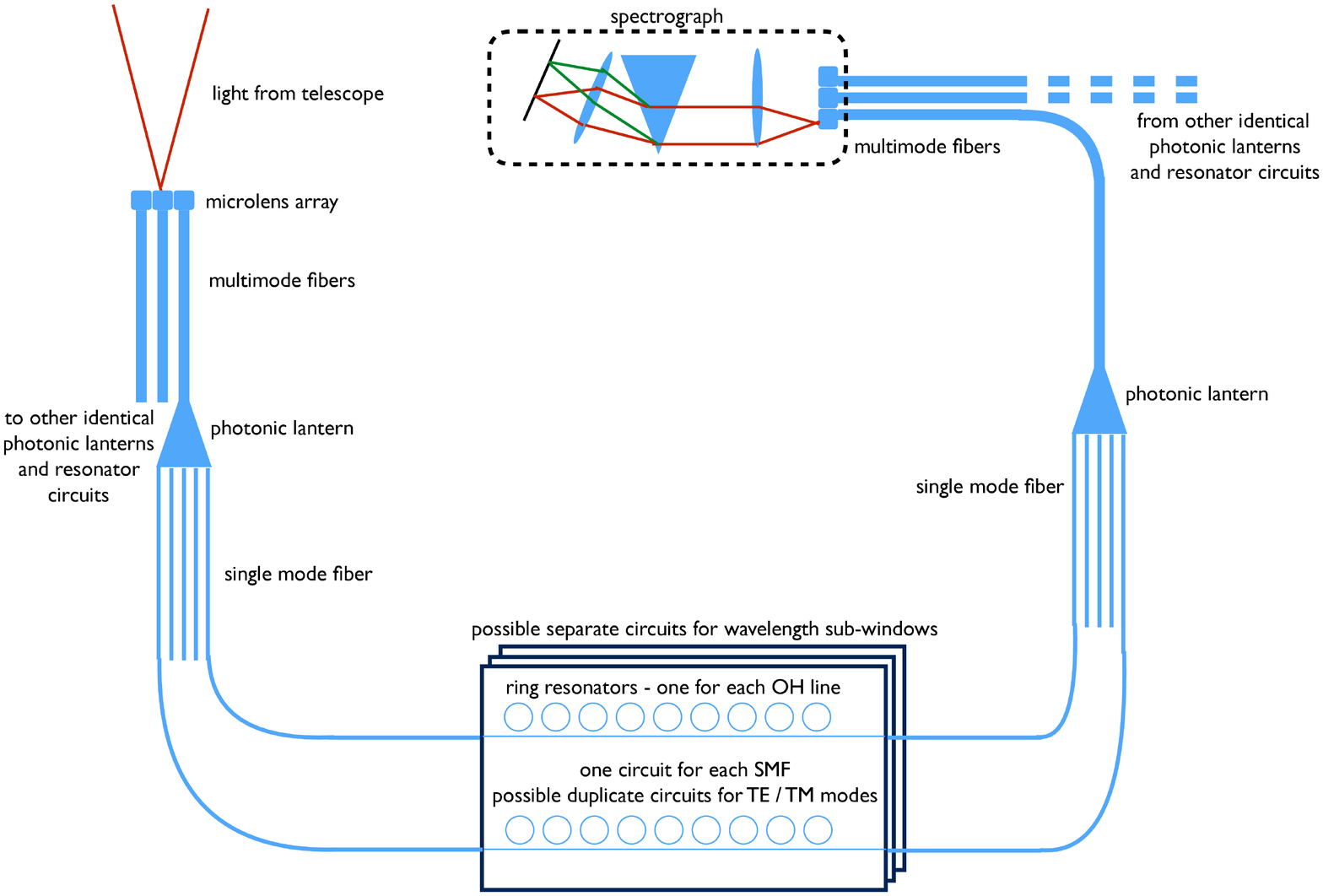}
\caption{Sketch of a ring resonator based OH suppression instrument.}
\label{fig:rrsketch}
\end{figure}

These considerations lead to  a design such as that sketched in Figure~\ref{fig:rrsketch}, which is very similar to the GNOSIS\cite{tri13a} and PRAXIS\cite{ell16} OH suppression FBG instruments.  Light from the telescope is focussed onto a microlens array, each element of which feeds a multimode fibre.  This stage is necessary to reduce the number of modes in each fibre, whilst enabling an adequate field of view, but may be superfluous if very highly multimode photonic lanterns are developed.   Each multimode fibre is converted into an array of single mode fibres via a photonic lantern.  Each single mode fibre is then coupled to an individual waveguide on a photonic circuit.  There can be multiple copies of each circuit, e.g.\ one for each multimode fibre.  If necessary there could be several different circuits for each multimode fibre, each covering a different wavelength range, the light being directed to the appropriate circuit by means of dichroic beam splitters.  

Each waveguide on the photonic circuit will couple to $N$ rings in series, where $N$ is the number of OH lines to be suppressed.  It may be necessary to split each waveguide into two, one for each of the TE/TM modes (see \S~\ref{sec:polar} below).

At the output of the waveguides the reverse processes will take place, with the single-mode fibres (SMFs) feeding individual multi-mode fibres (MMFs) via a photonic lantern, as well as recoupling any separate wavelength or polarisation tracks.   All the MMFs can then be aligned at the entrance to a spectrograph, allowing the spectrum of each microlens element to be measured, free of OH line emission. 

\subsection{Requirements of ring resonators for OH suppression}
\label{sec:ohreq}

\subsubsection{Number of rings, bandwidth and suppression factors}

Every notch in an OH suppression filter suppresses both the background and the signal from the object of interest.  Therefore, although more notches, deeper notches and wider notches will decrease the background they will also decrease the signal.  It is therefore desired to choose a combination of number, depth and width of notches to maximise the signal to noise.   We have calculated the increase in signal to noise as a function of number, depth and width of notches, assuming that the observations are sky background limited\footnote{N.B.\ If the sky background is strongly suppressed, and the interline continuum is very dark, then observations may be detector noise limited.}, i.e., the signal to noise is simply,
\begin{equation}
S/N \propto \frac{S}{\sqrt{B}}
\end{equation}
where $S$ is the integrated signal, and $B$ is the integrated background, over the passband of interest.  For the purposes of these calculations we assume idealised notches with a perfectly rectangular profile, and that all notches have the same depth and width (except where notches overlap).  Lines are selected to be suppressed in order of their average brightness.  Further we assume that the signal spectrum is flat; for many science cases there may be a particular feature which is important, rather than the whole spectrum.  Neverthless, these calculations give an idea of the general requirements for OH suppression.  We assume the sky-background to be as given in Ellis et al.\cite{ell08}, except with the interline continuum as measured by Maihara et al.\cite{mai93}.

\begin{figure}
\centering \includegraphics[scale=0.55]{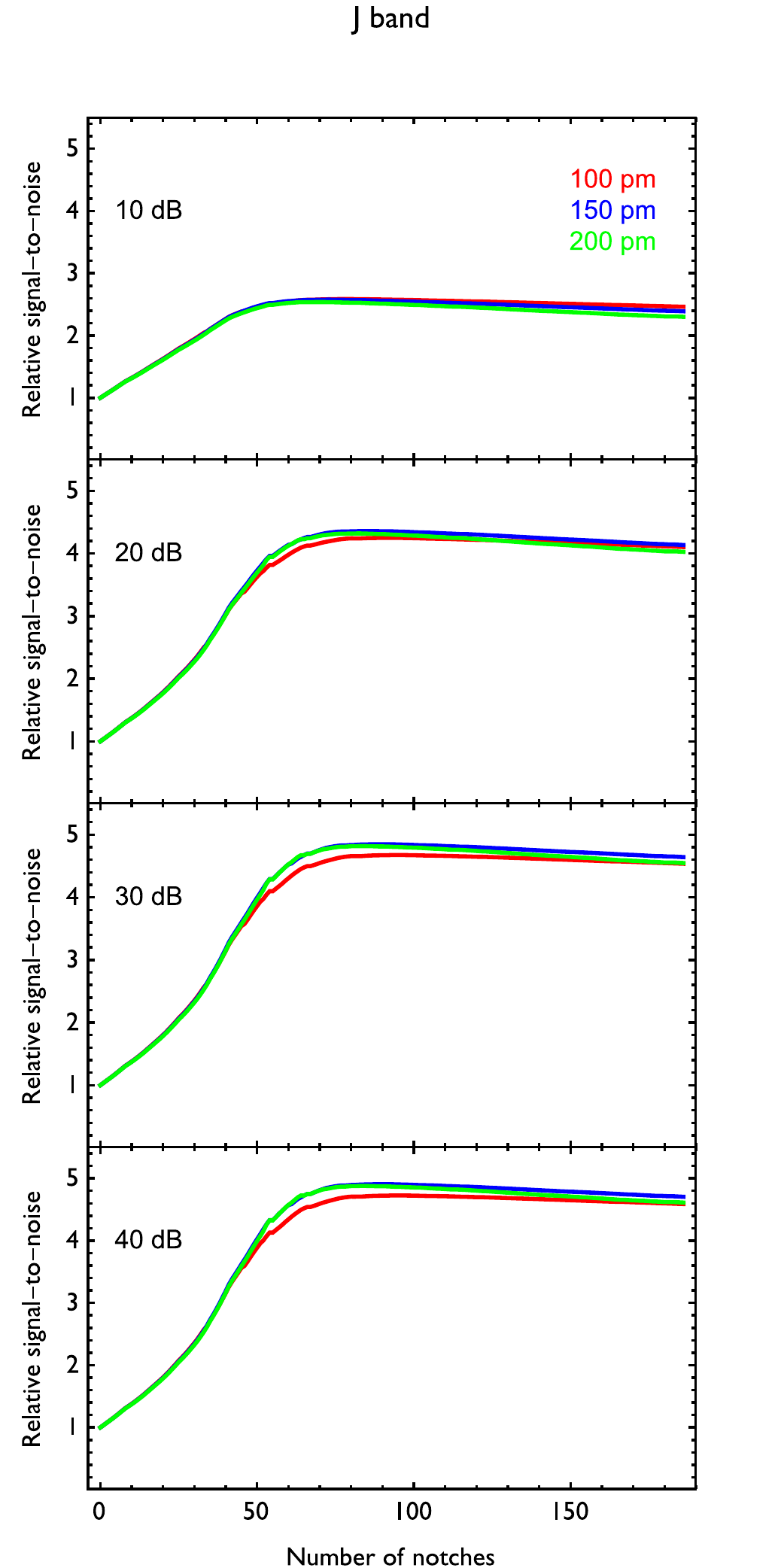}
\centering \includegraphics[scale=0.55]{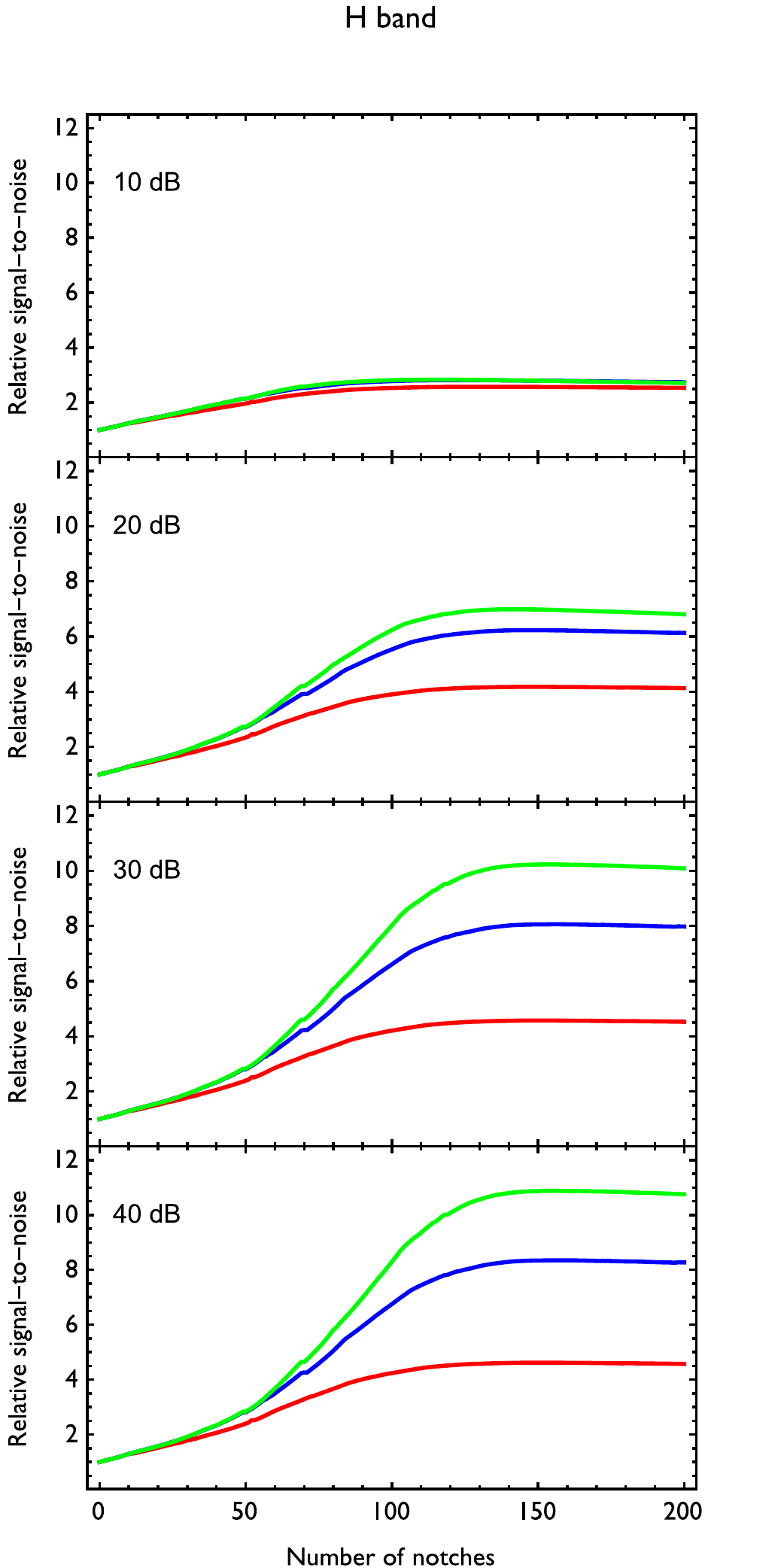}
\caption{The improvement in signal to noise due to OH suppression as a function of the number of notches, for notch widths of 100, 150 and 200 pm and for notch depths of 10, 20, 30 and 40 dB over the J (left) and H (right) bands.  The signal to noise is improved with deeper notches, up to $\approx 30$~dB, and with more notches up to $\approx 90$ notches in J and $\approx 150$ notches in H.  In the J band the notch width is not significant since the doublets are very closely spaced.  In the H band 200~pm wide notches are significantly better than narrower notches.}
\label{fig:nnotches}
\end{figure}

Figure~\ref{fig:nnotches} shows the improvement in signal to noise as a function of the number of notches, for notch widths of 100, 150 and 200 pm and for notch depths of 10, 20, 30 and 40 dB over the J and H bands.  The signal-to-noise improves with increasing notch depth, but the difference between 30 dB and 40 dB notches is very slight.  In the J band, the notch width is relatively unimportant between 100 and 200 pm.  In the H band, 200 pm notches are better than narrower notches in all cases, except for the 10 dB notches.  The optimal number of notches for each combination of notch width and depth are given in Table~\ref{tab:nnotches}.  In both the J and H band these correspond to approximately 1 notch for every 2~nm of passband, which is a useful approximation when considering shorter passbands.

The improvement in signal to noise will be compromised by the total throughput of the system.  However,  because $S/N \propto \sqrt{\eta}$, where $\eta$ is the throughput, the total end-to-end throughput of the system need only be $>4$~\% if the OH suppression increases the $S/N$ by a factor 5.  To be competitive with FBG OH suppression the total throughput of the OH suppression system itself should be $>50$~\%.

\begin{table}
\begin{center}
\caption{The optimal number of notches for maximum signal-to-noise for various notch widths and depths.}
\label{tab:nnotches}
\begin{tabular}{l|lll|lll}
& \multicolumn{6}{c}{Notch width (pm)}\\
& \multicolumn{3}{c}{J} & \multicolumn{3}{c}{H} \\
Notch depth (dB) & 100 & 150 & 200 & 100&150&200 \\ \hline
10& 78 & 73 & 71& 128 & 123 & 118 \\
20 & 93 & 84 & 82 & 147 & 147 & 141 \\
30 & 95 & 90 & 84 & 148 & 157 & 154 \\
40 & 95 & 90 & 84 & 148 & 157 & 156
\end{tabular}
\end{center}
\end{table}

Figure~\ref{fig:nnotches} suggests that the suppression factors should be $\sim 30$~dB, but after this there are diminishing returns.  The notch depth, $D$, is given by
\begin{equation}
\label{eqn:dsimp}
D_{\rm simple} = \frac{\left(\alpha - t\right)^{2}}{\left(\alpha t -1\right)^{2}},
\end{equation}
with the throughput of one passage around the ring given by $\alpha$, and the self-coupling along the input waveguide by $t$. 
Note that if $\alpha = t$ then the transmission is always zero on resonance  for a ring resonator with no drop port. Therefore to increase the notch depth $\alpha$ and $t$ should be made as equal as possible.   Other considerations, such as high $Q$ factor and high inter-notch transmission require $\alpha$ and $t$ to be close to 1; in this limit  the tolerances on $\alpha \approx t$ are rather tight,  since a slight inequality reduces the notch depth.
 
\subsubsection{Free spectral range}
\label{sec:fsr}

Because the OH lines are not periodic with frequency or wavelength, each individual line must be suppressed with an individual ring.  It could be possible to suppress two lines with the same ring if the FSR can be matched to the spacing between them without compromising the other requirements such as notch depth, notch profile and interline continuum. 

In general then, the FSR must be larger than the passband of interest.  The J and H band filters in the Mauna Kea filter set\cite{tok02}
are 160 and 290~nm wide respectively.  The free spectral range of a simple ring resonator, $\Delta \lambda$ is given by,
\begin{equation}
\label{eqn:fsr}
\Delta\lambda=\frac{\lambda^{2}}{L n_{\rm g}},
\end{equation}
where $n_{\rm g}$ is the group refractive index.

\begin{figure}
\centering
\includegraphics[scale=0.7]{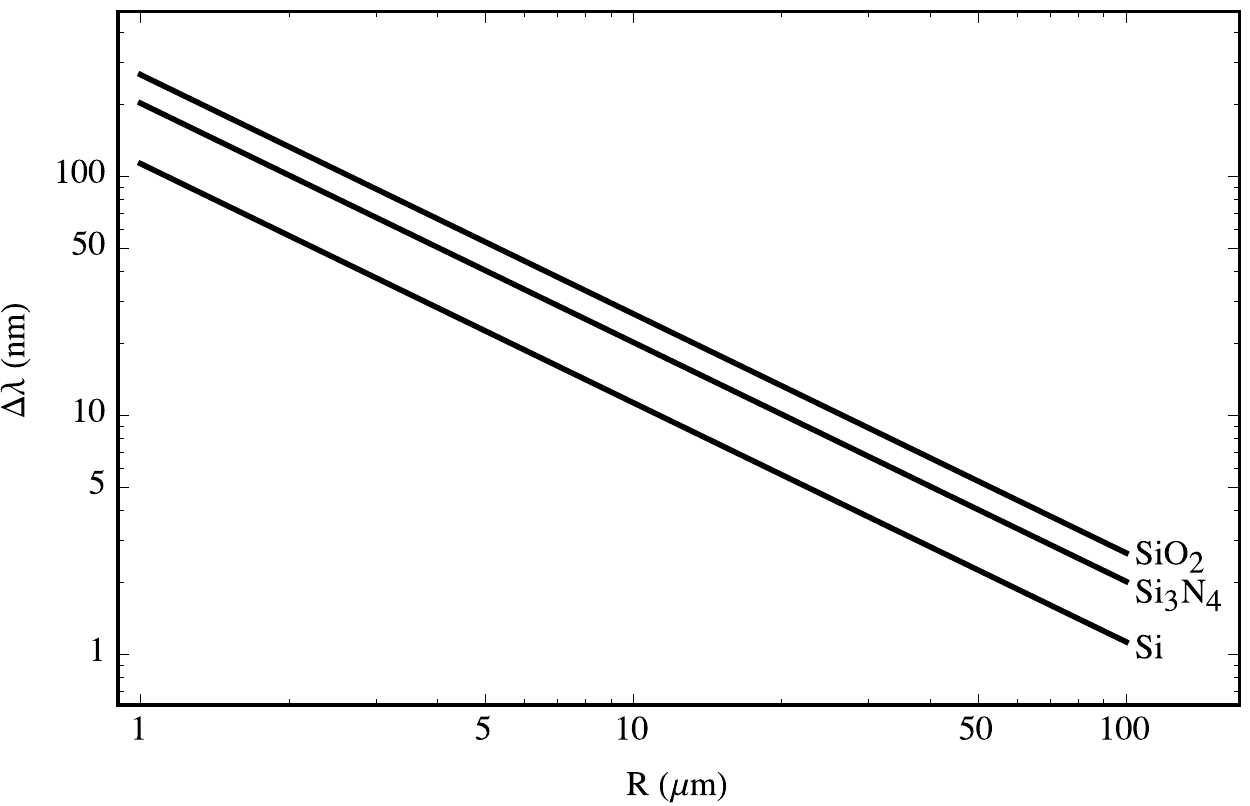}
\caption{The free spectral range at $\lambda = 1.55\ \mu$m of a circular ring resonator as a function of ring radius for Si$_{3}$N$_{4}$ and Si with $n_{\rm g} = 2.15$ and $n_{\rm g} = 3.85$ respectively, where the values come from our measured FSR (\S~\ref{sec:fsrresults}).}
\label{fig:fsr}
\end{figure}

Figure~\ref{fig:fsr} shows the free spectral range as a function of radius for a circular ring for the measured group indices of our Si and Si$_{3}$N$_{4}$ devices (\S~\ref{sec:fsrresults}).  
Comparing with these FSRs full coverage of J and H bands would require radii of 0.66 and 0.37~$\mu$m for  Si$_{3}$N$_{4}$, and Si respectively (the radii are approximately the same for each band).  These radii are too small to be made without significant bending losses.  To cover the entire J and H bands will therefore require several wavelength sub-windows of $30-70$~nm each.

\begin{figure}
\centering
\includegraphics[scale=0.26]{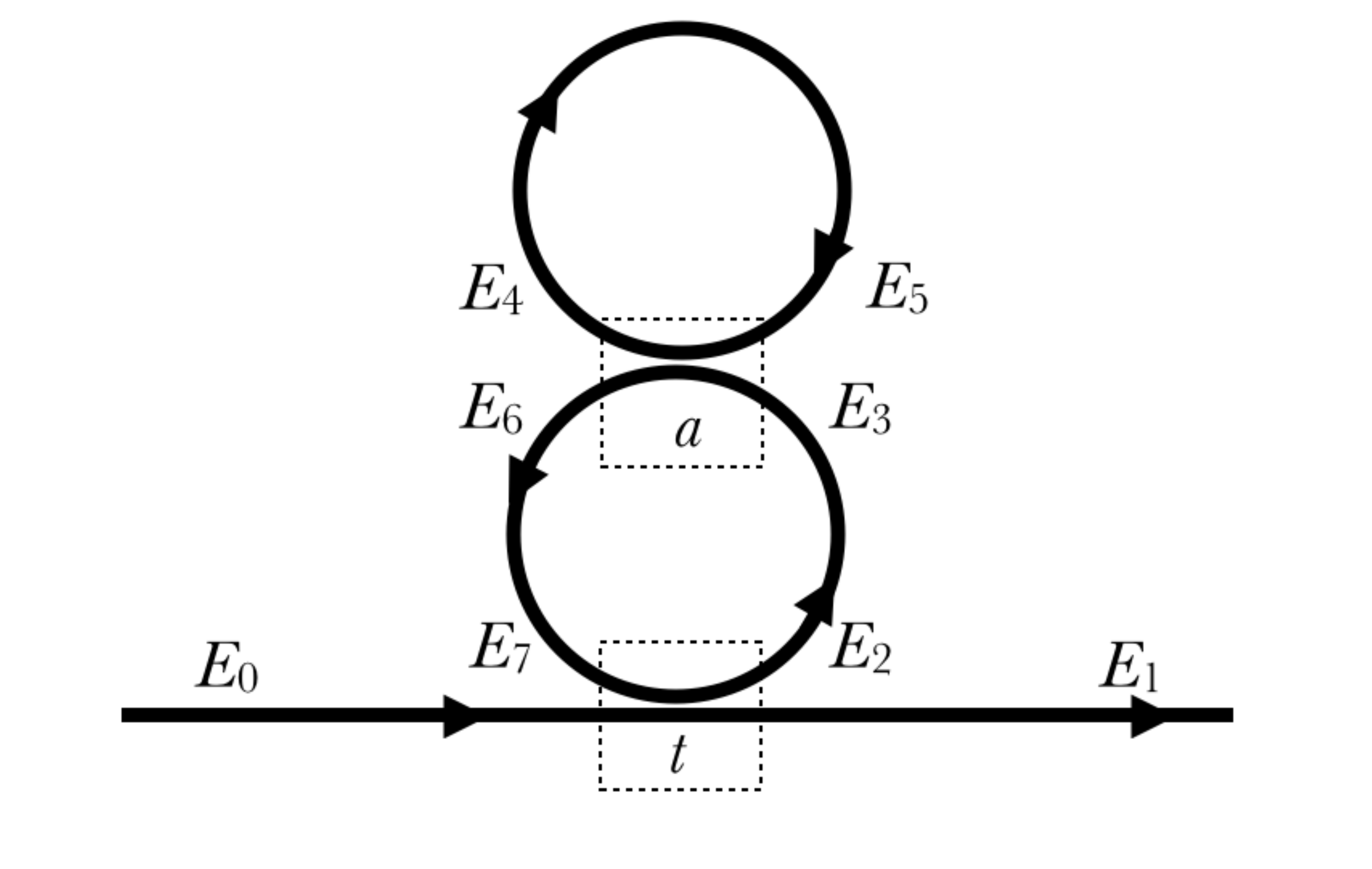}
\caption{Schematic diagram of the electric fields and coupling coefficients in a Vernier coupled resonator with no drop port.}
\label{fig:verniernotch}
\end{figure}

\begin{figure}
\centering \includegraphics[scale=0.7]{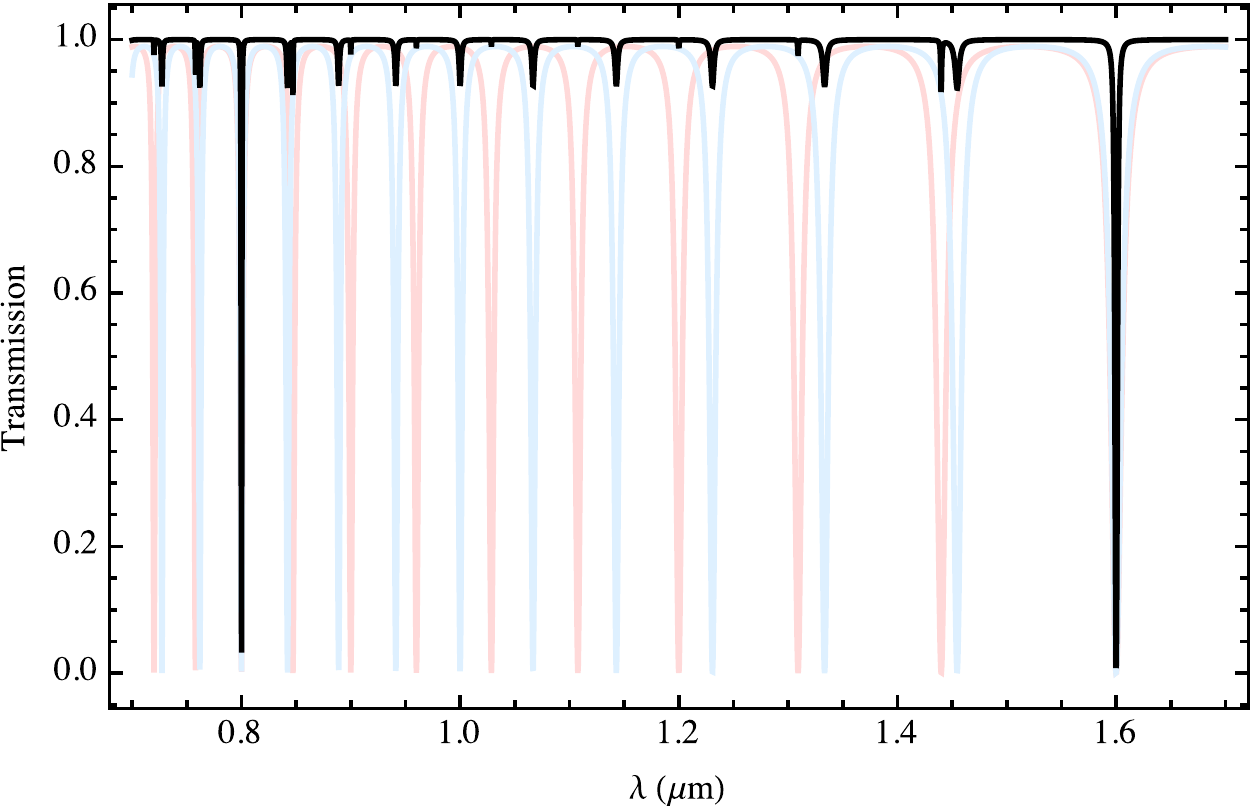}
\caption{The transmission function of a Vernier coupled double ring resonator (black) with $R_{1} = 9/10 R_{2}$,  and $\alpha = 0.998$, $\beta =0.98$ and $t = 0.9$, compared to devices made from the individual  rings (red and blue) with $\alpha=t=0.9$.}
\label{fig:verniersimple}
\end{figure}

It is possible to increase the FSR by using Vernier coupled ring resonators.  In this case, the two rings have a resonant wavelength in common, but have different radii, such that the neighbouring resonances from each ring do not overlap, see Figure~\ref{fig:verniernotch}.  For example if the circumference of the two rings a chosen to be $L_{1} =M \lambda/ n_{\rm e}$ and $L_{2} =P \lambda/ n_{\rm e}$,
then every $M$th notch of ring 1 will overlap with every $P$th notch of ring 2, with no overlapping in between, in which case the rings can be said to be Veriner coupled.    Figure~\ref{fig:verniersimple} shows an example transmission function with  $\lambda_{0}=1.6$, $M=9$ and $P=10$, compared to the transmission function of devices made from the individual  rings, in this example the FSR is increased tenfold.

\subsubsection{Inter-notch throughput}

The inter-notch transmission, $T$, 
is given by,
\begin{equation}
\label{eqn:tsimp}
T_{\rm simple} = \frac{\left(\alpha + t\right)^{2}}{\left(\alpha t +1 \right)^{2}};
\end{equation}
and 
is always higher for the same values of $\alpha$ and $t$.  The difference for optimally coupled Vernier coupled ring resonators with and without a drop port is insignificant.

\begin{figure}
\centering \includegraphics[scale=0.8]{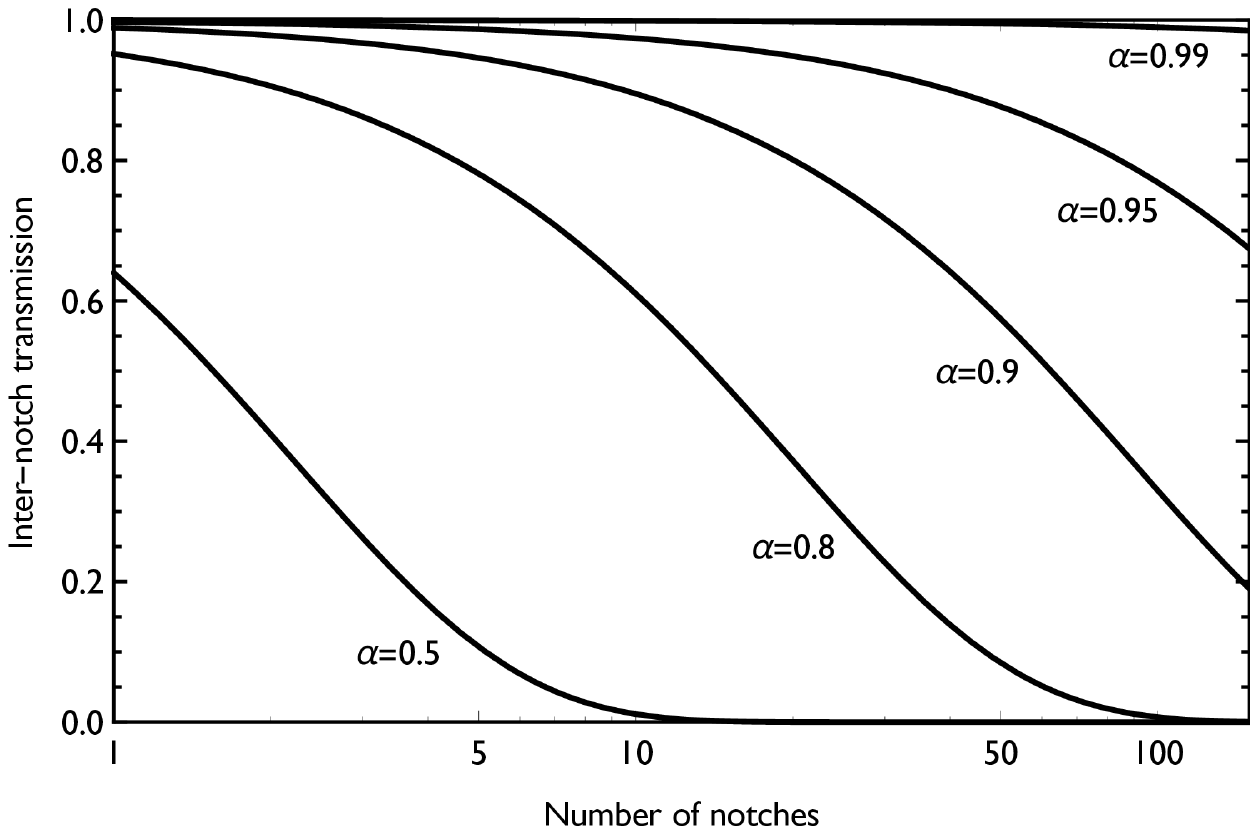}
\caption{The total transmission for multiple simple ring-resonators in series with $\alpha = t$ (or identically, simple Vernier coupled notch filters with $\alpha=\beta=t$)
as a function of the number of rings.}
\label{fig:tnotches}
\end{figure}

The total interline throughput will be given by $T^{N}$ where $T$ is the interline transmission of a single ring.   For example if using simple notch filters (equation~\ref{eqn:tsimp}), then we have,
\begin{equation}
\label{eqn:ttot}
T_{\rm tot} =  \left(\frac{\alpha + t}{\alpha t +1 }\right)^{2N},
\end{equation}
which is plotted for the case $\alpha = t$ for various $\alpha$ in Figure~\ref{fig:tnotches}.  

If it is required to have an inter-notch transmission of $\ge \tau$, then 
\begin{equation}
\alpha \ge \tau^{-\frac{1}{2N}}\left(1 - \sqrt{1 - \tau^{\frac{1}{N}}}\right),
\end{equation}
For example, 
an inter-notch transmission of $\tau=0.9$ is possible even for 100 rings in series if $\alpha=t>0.96$.

N.B.\ the requirements on high $Q$ for OH suppression,  usually require higher $\alpha$ and $t$ than the requirements on internotch throughput.  That is, if the rings have high $Q$, they will also have high inter-notch throughput.

\subsubsection{Coupling to astronomical instruments}
\label{sec:injection}

One of the major challenges in using ring resonators in astronomical instrumentation is feeding light from the telescope into the waveguides at high efficiency.  Indeed this is a problem for all astrophotonic instruments that use single mode waveguides.

This problem has two parts: (i) feeding the light from a telescope beam which has an $A\Omega$ etendue equivalent to a highly multimode waveguide into single mode fibres, (ii) feeding the light from a single mode fibre into the waveguides.

The first part of the problem can be solved either through the use of photonic lanterns\cite{leo05,bir16}, which split a multimode fibre into an array of SMFs at high efficiency, or through the use of extreme adaptive optics\cite{jov16,jov16b}.

The second part of the problem arises because ring resonators require high index contrast waveguides in order to achieve the small bending radii necessary for the large FSR without introducing unacceptable bending losses.  High index contrast waveguides are necessarily narrow to maintain single mode operation -- see section~\ref{sec:man}.  Therefore there is a large mismatch between the mode field diameter of the SMFs and the waveguides.  Straight butt-coupling of a SMF-28 fibre to a $\approx 400 \times 300$~nm Si waveguide
would lead to an insertion loss of $\approx -13$~dB, with an identical loss at the output coupling.

High efficiency fibre to chip coupling is an active area of research within the photonics community\cite{tho16}.  As such there are several promising solutions to this problem.

One solution is to use grating couplers, in which a diffraction grating is written into the surface of the waveguide, which thus diffracts light of a specific wavelength at specific angles. A fibre can therefore be placed at the corresponding angle to the waveguide and thereby collect the diffracted light from the waveguide, or in reverse, feed light into the waveguide.

Following such strategies grating couplers can reach peak efficiencies of $> -1$~dB. However, these efficiencies peak in a narrow range of wavelengths. Typical 3 dB bandwidths are $\approx 40$ -- $70$~nm. Compared to the required astronomical passbands of  $\sim 160$ -- 290~nm (\S~\ref{sec:ohreq}), grating couplers are very narrow band. They could find applications for particular niche science cases in which only specific wavelengths are to be targeted.

\begin{figure}
\subfigure[Inverted taper sketch]{
\centering \includegraphics[scale=0.25]{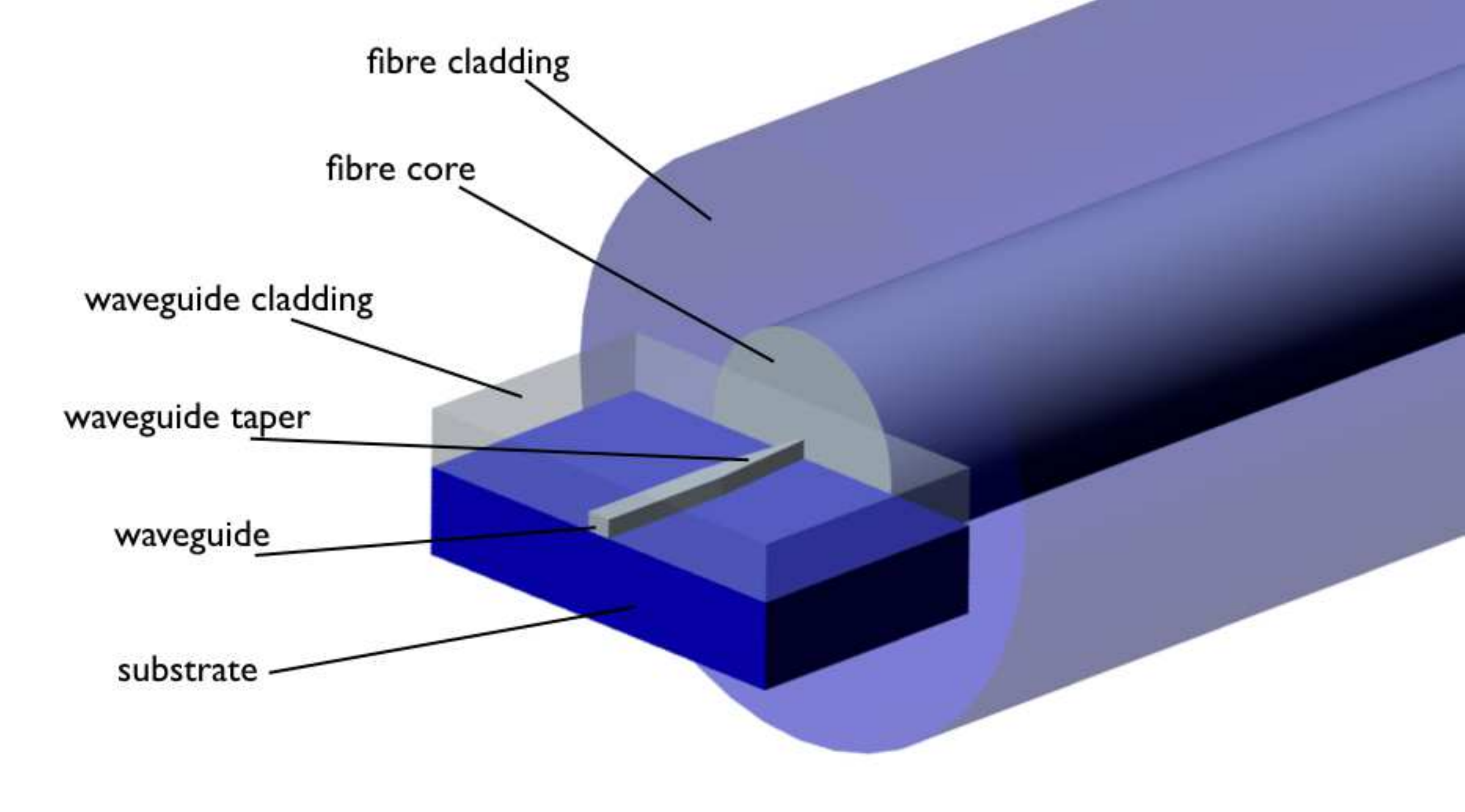}
}
\subfigure[Inverted taper with a cantilever and V-groove]{
\centering \includegraphics[scale=0.27]{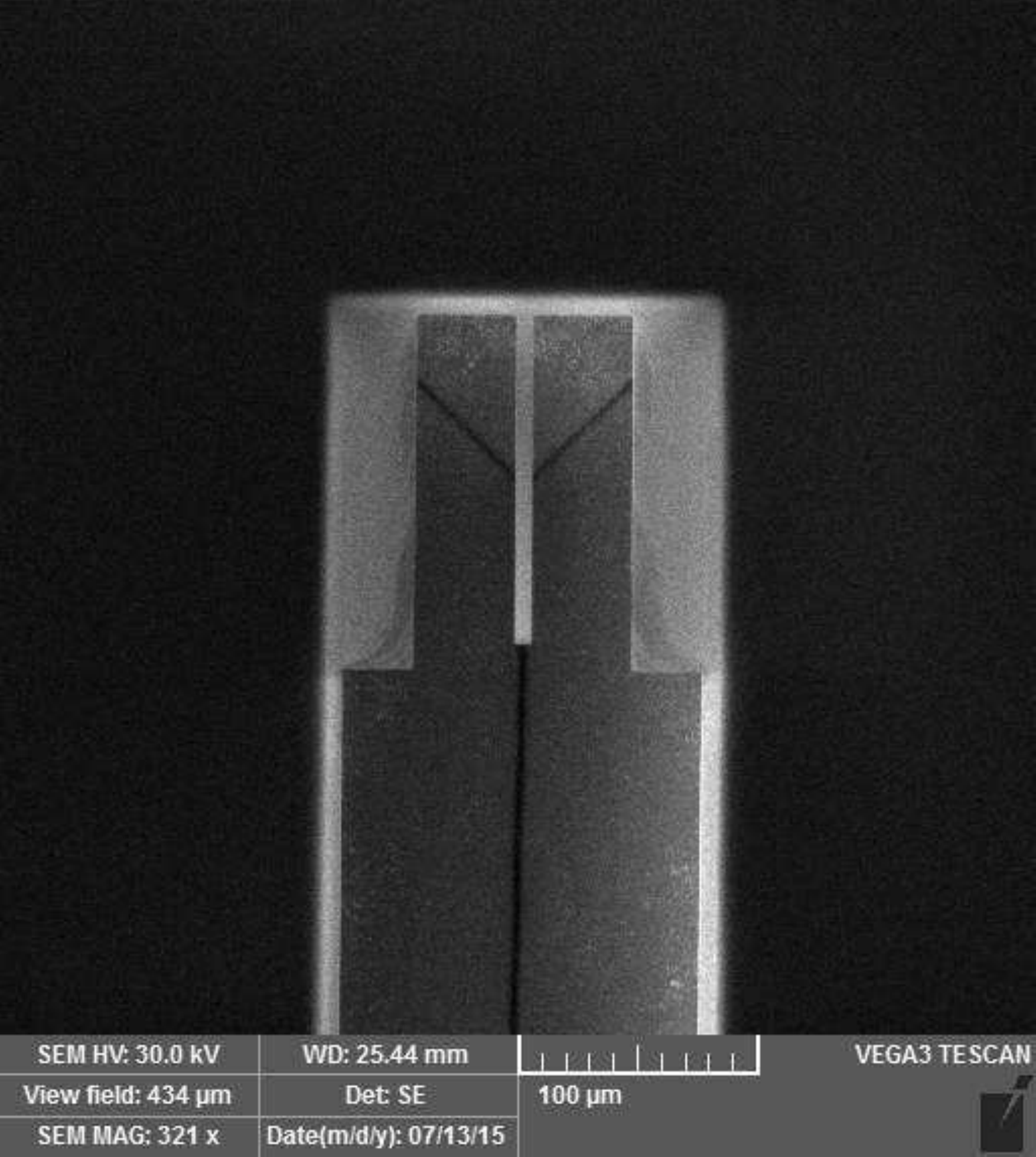}
}
\caption{(a.) Sketch of an inverted taper butt-coupled to a fibre.  The dimensions are not drawn to scale, in order to make all components visible. (b.) An SEM photograph of one of our tested cantilever devices based on the work of Galan\cite{gal10}.}
\label{fig:invtaper}
\end{figure}

\begin{figure}
\centering \includegraphics[scale=0.8]{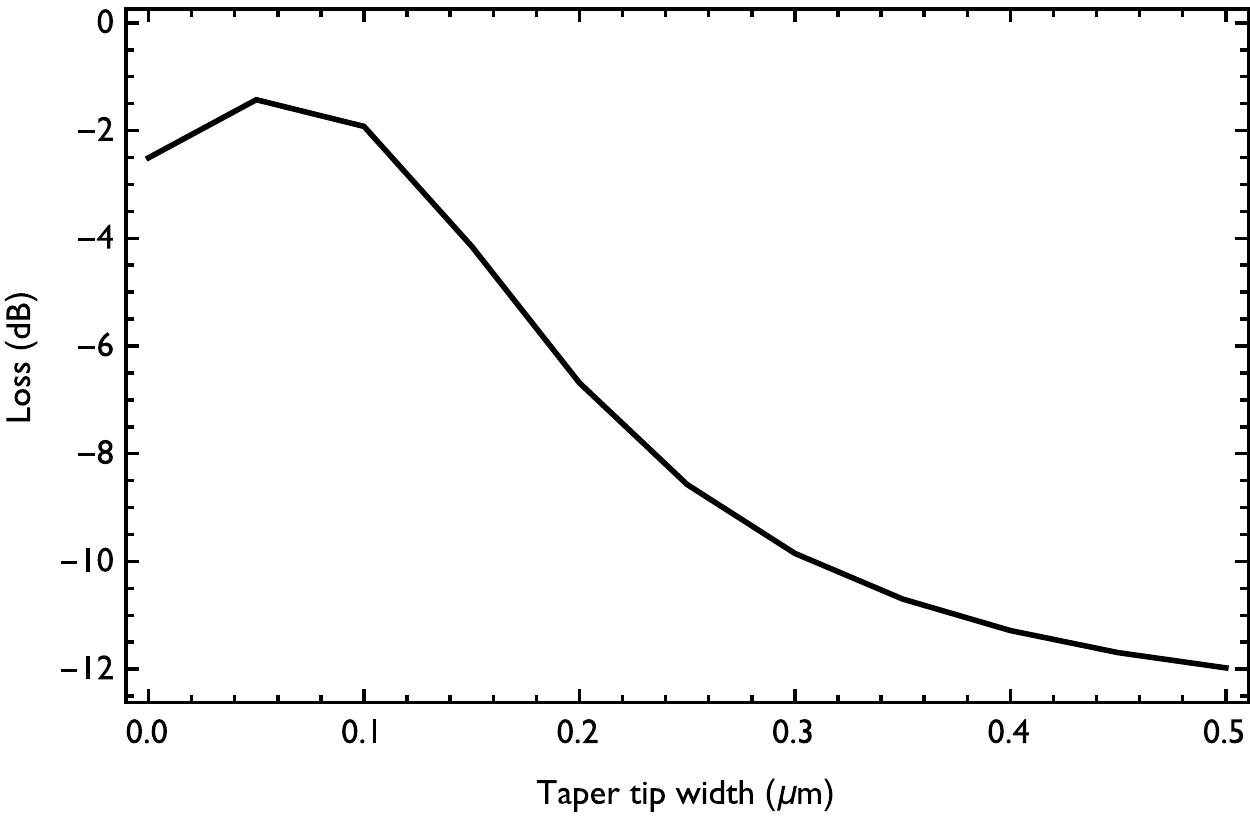}
\caption{The power coupled between a perfectly aligned 8~$\mu$m core fibre, and a Si$_{3}$N$_{4}$ waveguide in SiO$_2$ cladding, of  height 650 nm and width 900 nm, which tapers over a distance 400~$\mu$m to a final tip width as plotted.}
\label{fig:coupledpower}
\end{figure}

A more promising solution for astronomical requirements is to use an inverted taper on the waveguide, whereby the waveguide decreases in width at the edges of the chip -- see Figure~\ref{fig:invtaper}a. As the width of the waveguide narrows, more of the light is squeezed out of the core into the evanescent field, increasing the mode field diameter. Note that the waveguide need only be tapered in one dimension, viz.\ the width, in order to increase the mode field diameter in this way. 
Insertion losses of less than 10\% have been demonstrated in multiple publications \cite{taper_IBM,taper_Cornell,taper_China,taper_Harvard}. 
Figure~\ref{fig:coupledpower} shows the results from {\sc RSOFT BeamPROP} simulations of the coupled power between an 8~$\mu$m core fibre, and a Si$_{3}$N$_{4}$ waveguide in SiO$_{2}$ cladding, of height 650~nm and width 900~nm, which tapers over a distance 400~$\mu$m to a final tip width as plotted. The maximum coupled power of $\approx -1.6$~dB occurs when the waveguide is tapered down to $\approx 50$~nm.

Coupling between a fibre and an inverted waveguide taper may be further improved in several ways. For example, the tapered waveguide may be covered in a polymer cladding which acts as a new waveguide better matched to the fibre, and eases the alignment tolerances; the fibre itself may also be lensed or of a high NA in order to match the fibre mode to that of the waveguide.

An interesting refinement of inverted tapers was suggested by Galan\cite{gal10}.  In this method the inverted taper sits on top of a `cantilever' type structure, which is etched from the SiO$_{2}$ substrate, as in Figure~\ref{fig:invtaper}b.  Thus, as the light is squeezed out of the waveguide during the taper, it is guided by the SiO$_{2}$ cantilever, which now acts as a waveguide surrounded by air.  This cantilever waveguide has an MFD much better matched to a SMF, increasing the coupling, and easing the alignment tolerances.  Furthermore,  a V-groove can be etched under the cantilever to facilitate alignment. However, the V-grooves/cantilever system, with silicon waveguides, proved challenging to fabricate and assemble without damaging the waveguides and cantilevers.   As a result, we decided to focus our current fabrication efforts on the study of the ring resonators with more conventional butt-coupling schemes.

\subsubsection{Quality factor}

The quality of a ring resonator is equal to its resolving power, $\lambda/\Delta \lambda$.  For OH suppression notch widths of $\Delta \lambda \approx 100$ -- $200$~pm are necessary, which are equivalent to Q factors of 7750 -- 15500 at $\lambda = 1.55$~$\mu$m.

The Q factor can be increased by increasing  the self-coupling factor $t$ through careful control of the gap size, and the throughput values $\alpha$ through minimising leaking into the substrate with a thicker SiO$_{2}$ cladding above and below the waveguides.
Note that high $Q$ devices are certainly feasible\cite{bae04,xu08}.

\subsubsection{Notch wavelengths and shape}
\label{notchspec}

The wavelengths of the notches need to be aligned to the OH lines.  The accuracy of this alignment needs to be  better than half of the notch width, such that the notch will always cover the OH lines.  Figure~\ref{fig:nnotches} shows the best improvement in signal to noise is achieved with notches 200~pm wide.  The H band signal to noise is more sensitive to the notch width than the J band, because the average OH doublet spacing is larger in the H band than in J.  In either case the notch wavelengths must be tuned to within $\approx 100$~pm of the OH line wavelengths.

Ideally, the notches should be as rectangular as possible.  This ensures that the maximum suppression occurs across the full width of the notch, allows for slight misalignments in wavelength, and avoids suppressing the interline regions.

\begin{figure}
\centering \includegraphics[scale=0.6]{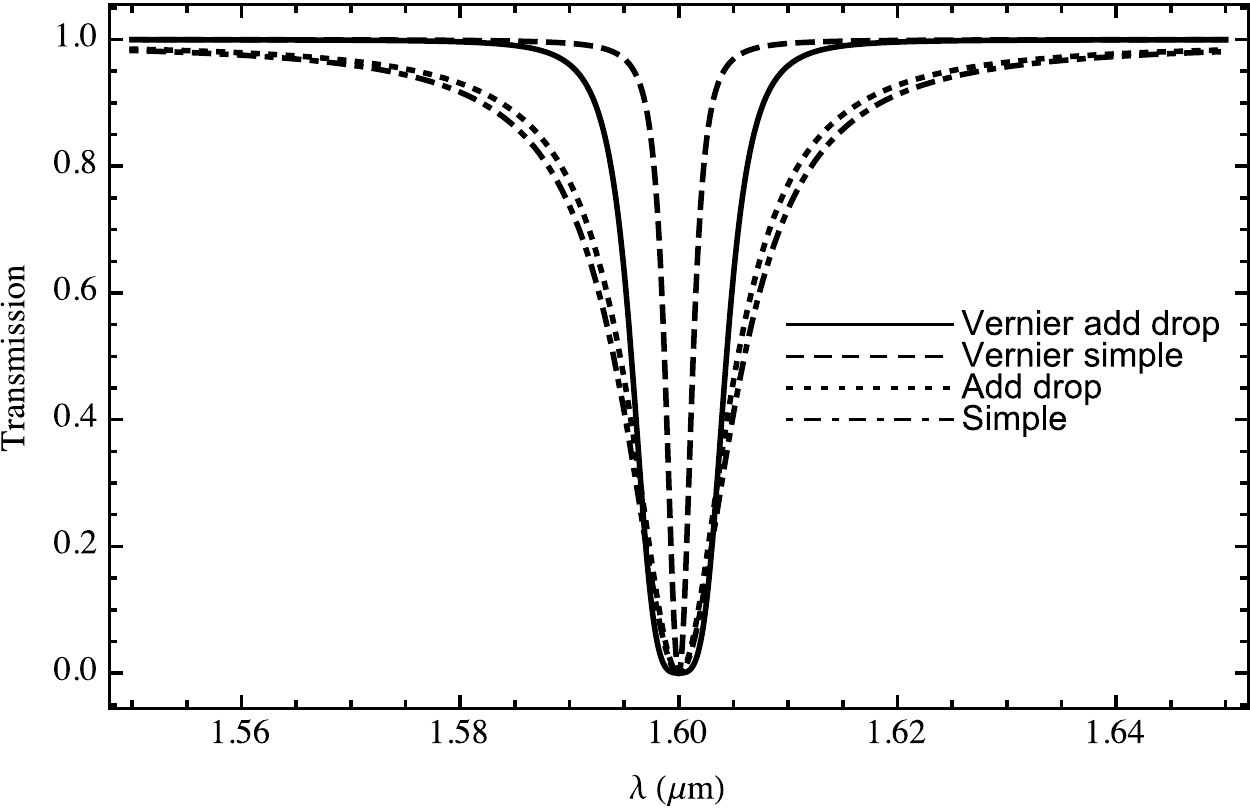}
\caption{Comparison of the notches of optimised ring-resonators consisting of a single ring with no drop port ($\alpha=t=0.9$), a single ring with a drop port ($\alpha=1$, $t=0.9$), a double ring with no drop port ( $\alpha=0.998$, $\beta=0.98$, $t=0.9$), and a double ring with a drop port ($\alpha=0.998$, $\beta=0.98$, $t=0.9$).}
\label{fig:notchcf}
\end{figure}

In practice the notch shape is  determined by the  layout of the device, i.e., the number of rings and the presence or absence of a drop port.  Figure~\ref{fig:notchcf} shows that Vernier coupled rings are squarer than single rings, and indeed more rings would make the notch even squarer\cite{xia07}.

\subsubsection{Polarisation}
\label{sec:polar}

In general rectangular waveguides have two modes of polarisation which are approximately transverse electric (TE) or transverse magnetic (TM).  Each polarisation mode will have a different effective index, and therefore a different set of resonant wavelengths.

Therefore, we may require double sets of ring resonator circuits, one for each polarisation, or polarisation independent waveguides.  In this case the light must be split into orthogonal polarisations, either on or off chip, and in a wavelength independent manner over the wavelength range of interest.

However, it has been shown that it is  possible to  tune the waveguide geometry precisely to enable approximately polarisation independent behaviour, even for small waveguide structure and ring-resonators as small as $R=3$~\um\cite{ang09}.  Further work is necessary to see if this is practicable whilst meeting the other requirements for OH suppression, especially maintaining polarisation independence of wavelength  ranges of $30$~nm or more.

\subsubsection{Stability}\label{sec:stability}

Ring resonators are used as temperature sensors, making use of the fact that the resonant wavelength will shift with temperature due to changes in the refractive index of the materials used and the dimensions of the waveguide, which will both change the effective index of the waveguides, and also the changes of the ring radii, which will directly affect the resonant wavelength.

These effects need to be minimised via temperature control in order to use ring-resonators for astronomy, when it is imperative that the resonant wavelengths stay aligned with the OH lines.  Fortunately, since ring resonators are very small the temperature control is not restrictive.

\begin{figure}
\centering \includegraphics[scale=0.7]{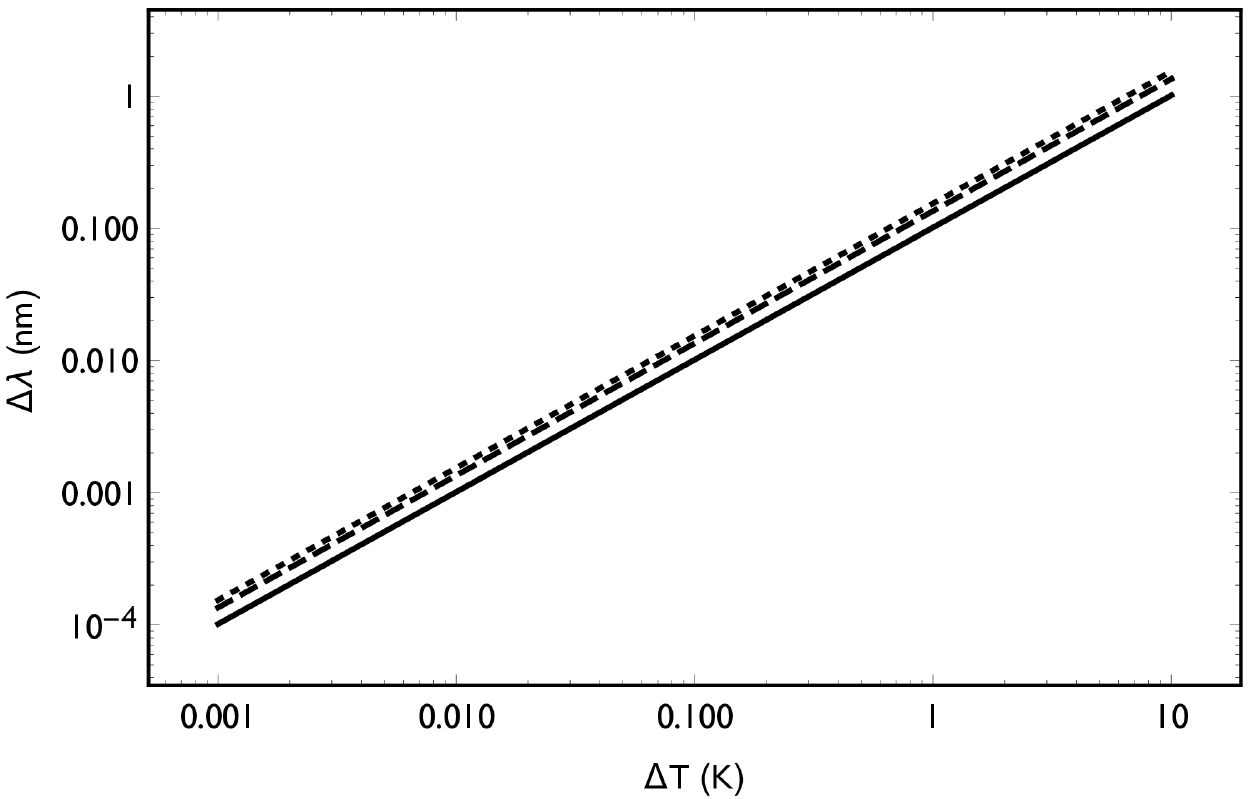}
\caption{The change in resonant wavelength as a function of temperature for a SOI waveguide with $R=5$~$\mu$m, at $\lambda = $ 1.1, 1.55 and 1.8~$\mu$m (black, dashed, dotted lines respectively).}
\label{fig:tempdep}
\end{figure}

Rouger, Chrostowski, \& Vafaei (2010)\cite{rou10} have determined the magnitude of these temperature effects for SOI waveguides.  The changes in dimensions due to thermal expansion are negligible, with only $0.038\%$ increase in the dimensions for a temperature change of 107~K.  We have used their changes in effective index  to calculate the shift in resonant wavelength as a function of temperature and wavelength for a SOI waveguide with width 620~nm and height 250~nm, see Figure~\ref{fig:tempdep}.  Controlling the temperature to within $0.3$~deg will stabilise the resonant wavelength to within 50~pm, such  that even a 100~pm wide notch will remain aligned to the OH lines.

Temperature stabilised ring resonators have been demonstrated\cite{lee12} using a resistive thermal device to provide temperature feedback to a 
thermo-electric cooler. A wavelength stability of 1 pm was achieved over a 24hr period for a single peak.

\subsubsection{Summary of requirements}
\label{sec:ohreqsum}

In the next section we will discuss the practical aspects of implementing  devices satisfying these requirements, but already there are some theoretical limitations to realistic devices.

In summary, for efficient OH suppression we require a total of $\approx 90$ notches with a width of $\approx 150$~pm and a depth of $\approx 30$~dB for optimal suppression of the J band, and $\approx 150$ notches with a width of $\approx 200$~pm and a depth of $\approx 30$~dB for optimal suppression of the H band; Table~\ref{tab:nnotches}.  The total throughput of the OH suppression system should be $>50$~\% to be competitive with other schemes.

However, the total passbands of the J and H bands would require FSRs corresponding to radii $< 1$~$\mu$m, which are would be extremely lossy.
Thus the J and H passbands must be broken up into a series of smaller bands, each of which is suppressed separately (Fig.~\ref{fig:rrsketch}).  For example, a ring with radius $6.5$~$\mu$m would have 
$\alpha\approx 0.9$ and 
a FSR of $\approx 31$~nm at $\lambda=1.6$~$\mu$m.  A 31~nm bandpass would require $\sim 15$ notches (Table~\ref{tab:nnotches}).   Fifteen notches with $\alpha = t = 0.9$ would have a total interline throughput of $\approx 0.85$ (equation~\ref{eqn:ttot}).

\begin{figure}
\centering
\includegraphics[width=0.4\textwidth]{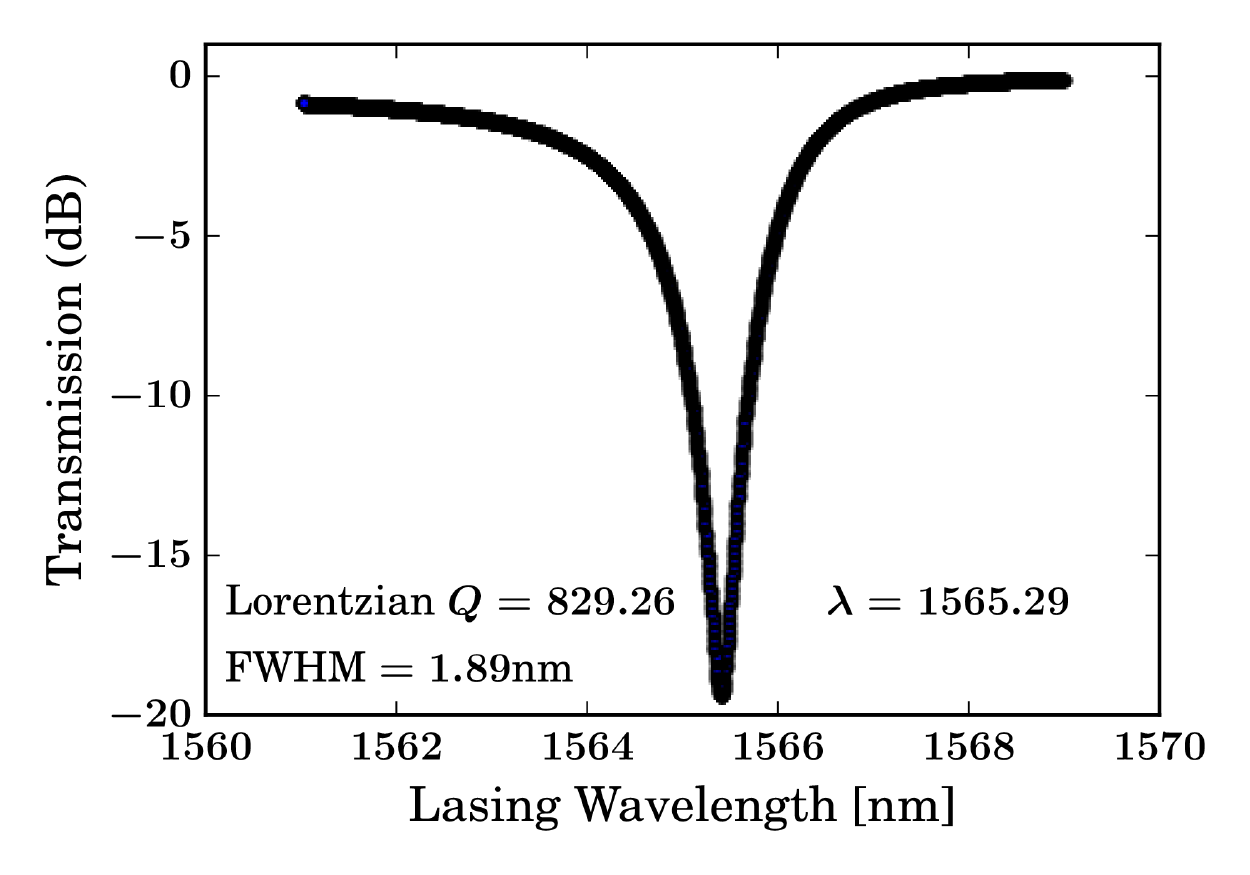}
\includegraphics[width=0.4\textwidth]{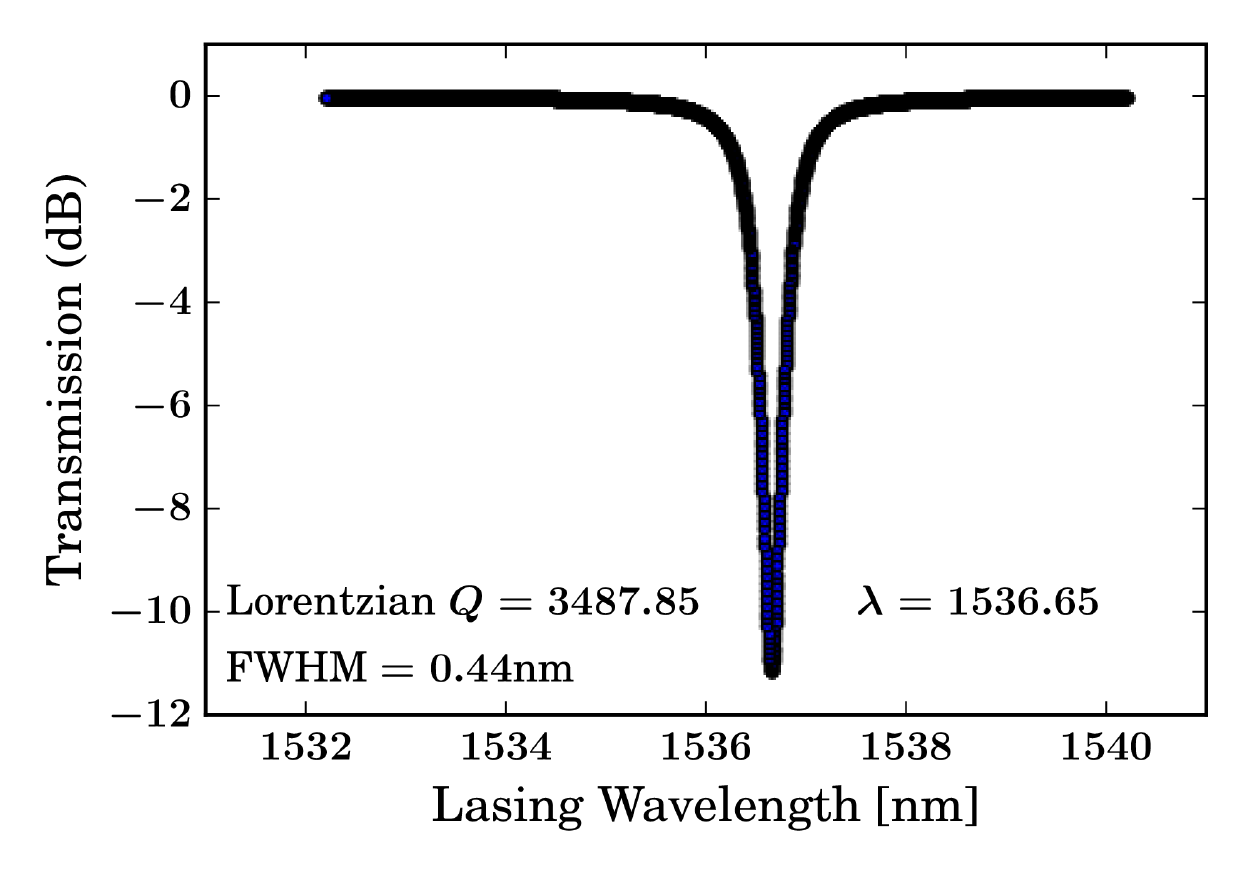}
\caption{(Left.)  The simulated response of a 5~\um\ Si$_{3}$N$_{4}$ ring,  with the parameters from a fit to a Lorentzian distribution.  This device has a FSR of 38nm.  (Right.) The simulated response of a 1.5~\um\ Si  ring has a higher $Q$ and a larger FSR of 73~nm.}
\label{fig:sims}
\end{figure}

However, to obtain a sufficiently high $Q$ factor would require even higher $\alpha$ and $t$.  FDTD simulations made with Lumerical show that to meet the requirements of high $Q$ and large FSR simultaneously with a single ring will require using Si waveguides, see Figure~\ref{fig:sims}.
Nine such `circuits' could cover the entire H band, noting that each would have to be reproduced for every spatial and polarisation mode (but given the extremely small dimensions of such devices, several circuits could fit on one wafer).  If Vernier coupled rings can be implemented then wider passbands and fewer individual circuits can be used.

\section{Practical implementation of ring resonators for OH suppression}
\label{sec:prac}

The requirements of ring resonators for OH suppression has been discussed in section~\ref{sec:ohsupp}.  We now turn our attention to the practical requirements of implementing ring resonators for OH suppression, taking into account the measured performance of real devices.  In particular we will discuss our own prototype devices, fabricated at the Center for Nanoscale Materials at Argonne National Laboratory (\S~\ref{sec:man}), and our testing of these prototypes, both at Argonne National Laboratory and at the Australian National Fabrication Facility at Macquarie University (\S~\ref{sec:lab}).

The main aim of our prototype manufacture and testing is to determine the correct design parameters necessary to meet the requirements outlined in section~\ref{sec:ohreq}.  For example,  it is necessary to determine the correct waveguide material, waveguide dimensions, cladding thickness, ring shape (circular or racetrack), gap size, and fibre-chip coupling method, to produce the correct wavelength and depth of suppression, FSR, Q factor, $\alpha$ and coupling parameters, and insertion losses,
such that we may ultimately develop devices containing many rings and tracks for OH suppression.

\subsection{Design considerations}
\label{sec:design}

Ring resonator literature provides considerable guidance in our design considerations, and many of the requirements outlined in section~\ref{sec:ohreq} have been individually met in published results.  For example, suppression of 43dB was achieved in \cite{ultra_suppression} with careful tuning of waveguide-to-ring couplings. A free spectral range of 63nm was achieved in \cite{Si_ring_HP}, as well as good suppression and Q factor.  As mentioned earlier, insertion losses of less than 10\% have been demonstrated in multiple publications \cite{taper_IBM,taper_Cornell,taper_China,taper_Harvard}.  Given the published results on suppression, FSR, and fibre-to-chip coupling, our focus in fabrication thus far has been to survey a variety of geometries including rings and racetracks, gap sizes, waveguide sizes, waveguide/ring materials, etc. to test our simulations as well as published results.   As discussed below in section~\ref{sec:doubleringresults}, we also have placed an emphasis on producing double rings on the same waveguide, to study the notch placement accuracy, an area not studied before in detail.

\subsection{Fabrication steps}
\label{sec:man}

The fabrication of the ring resonators has been carried out at Argonne's Center for Nanoscale Materials,  with an array of lithography, etching, and metrology tools.  The process is currently identical for silicon nitride and silicon wafers,  with the seven steps outlined below.

\begin{enumerate}
\item{Start with a three-layer wafer with either silicon or silicon nitride device layer, 1-3$\mu$m SiO2 dielectric, and Si substrate}
\item{Oxygen plasma to improve photo-resist adhesion}
\item{Spin-coat HSQ photo-resist 2000rpm for 60s, then bake at 150$^{\circ}$ C for 3 minutes}
\item{Expose with the JEOL JBX-9300FS electron beam lithography system, and develop the photo-resist with MF CD-26 for 2 minutes}
\item{Plasma etching of the SiN or Si device layer}
\item{Measure rings, gaps, and waveguides with a scanning electron microscope}
\item{Plasma-enhanced vapor deposition of SiO2 cladding}
\end{enumerate}

The entire process takes about six hours per wafer, and our most recent designs have more than ten waveguides on a 25mm square wafer. 

\begin{figure}[ht]
\centering \includegraphics[width=0.361\textwidth]{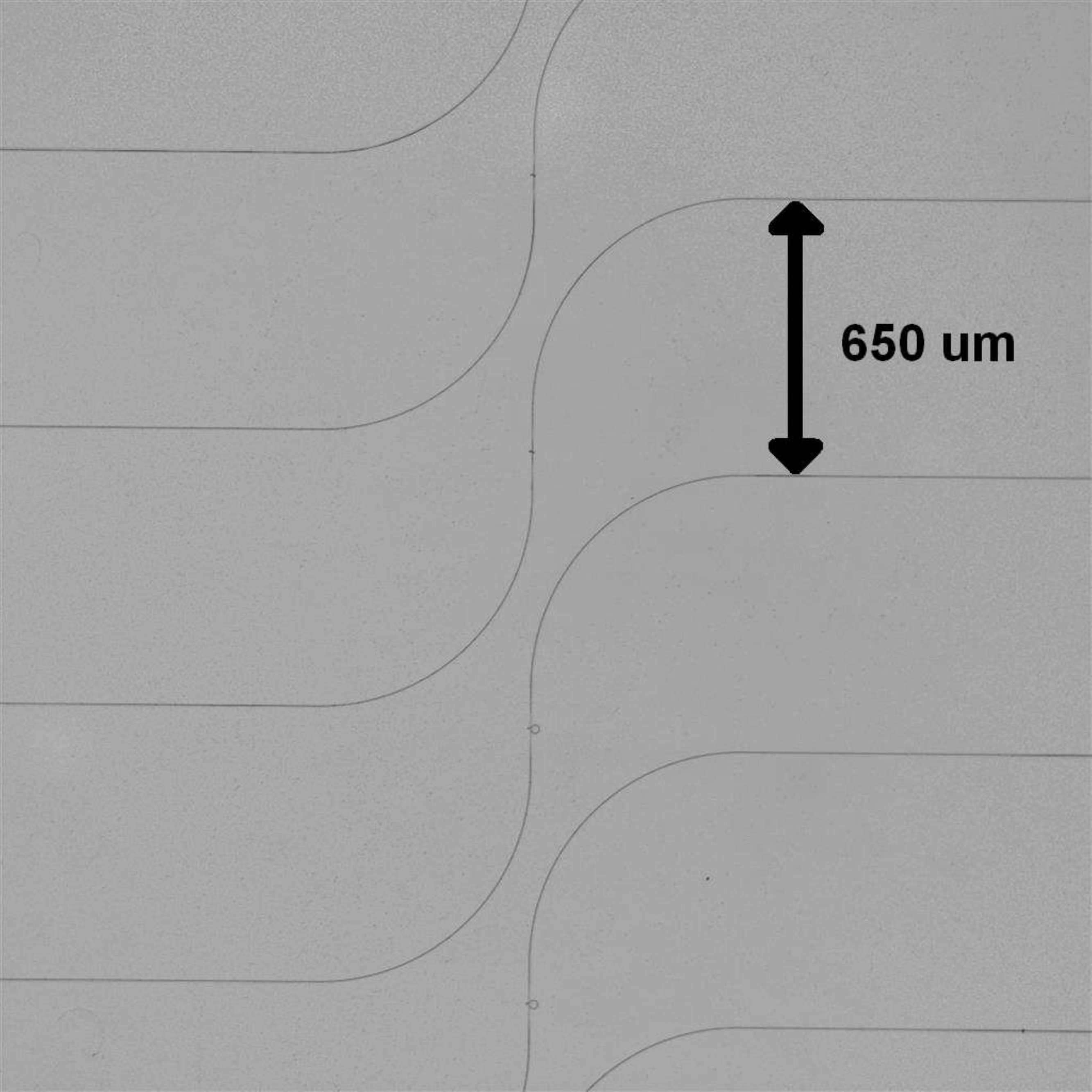}
\centering \includegraphics[width=0.4\textwidth]{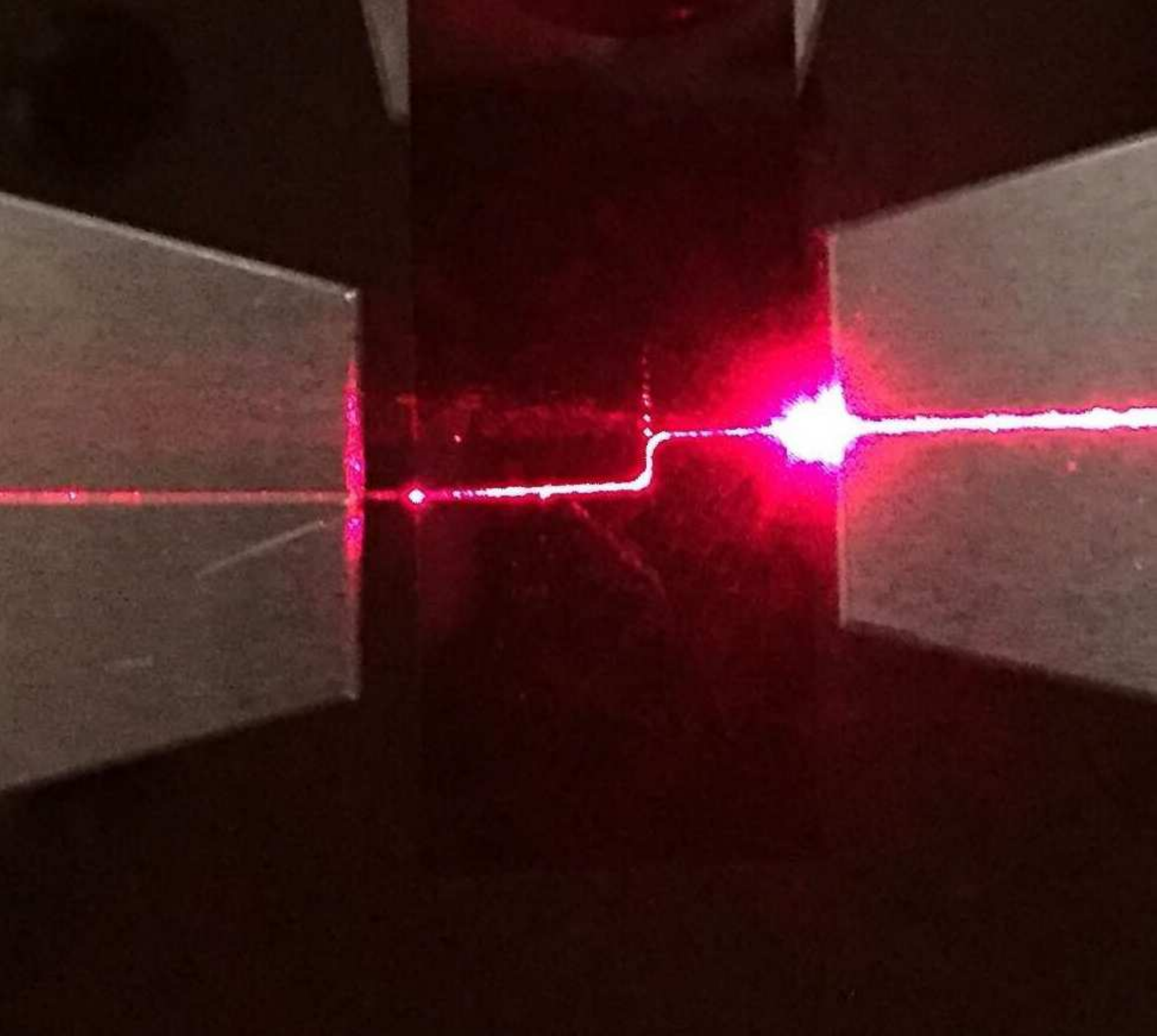}
\caption{(Left.) Microscope view of a multiple-waveguide wafer,  showing the fabricated waveguides in a s-shape.  The s-shape design reduces backgrounds from light outside of the waveguide traveling straight to the output fibre. (Right.) A zoomed-in view of fibres precisely aligned with the s-shaped waveguide on a silicon nitride wafer. Red laser light is sent into the input fibre on the right side, couples to the waveguide, and then proceeds through the output fibre on the left side.}
\end{figure}

\subsection{Test procedures and results}
\label{sec:lab}

Laboratory tests have been carried out with both Si and Si$_{3}$N$_{4}$ waveguides; Table~\ref{tab:tested} lists the number of devices tested so far, giving the material and basic design layout.  Tests have been carried out concurrently at Argonne National Laboratory and at the Australian National Fabrication Facility at Macquarie University.

\begin{figure}[ht]
\centering \includegraphics[width=0.5\textwidth]{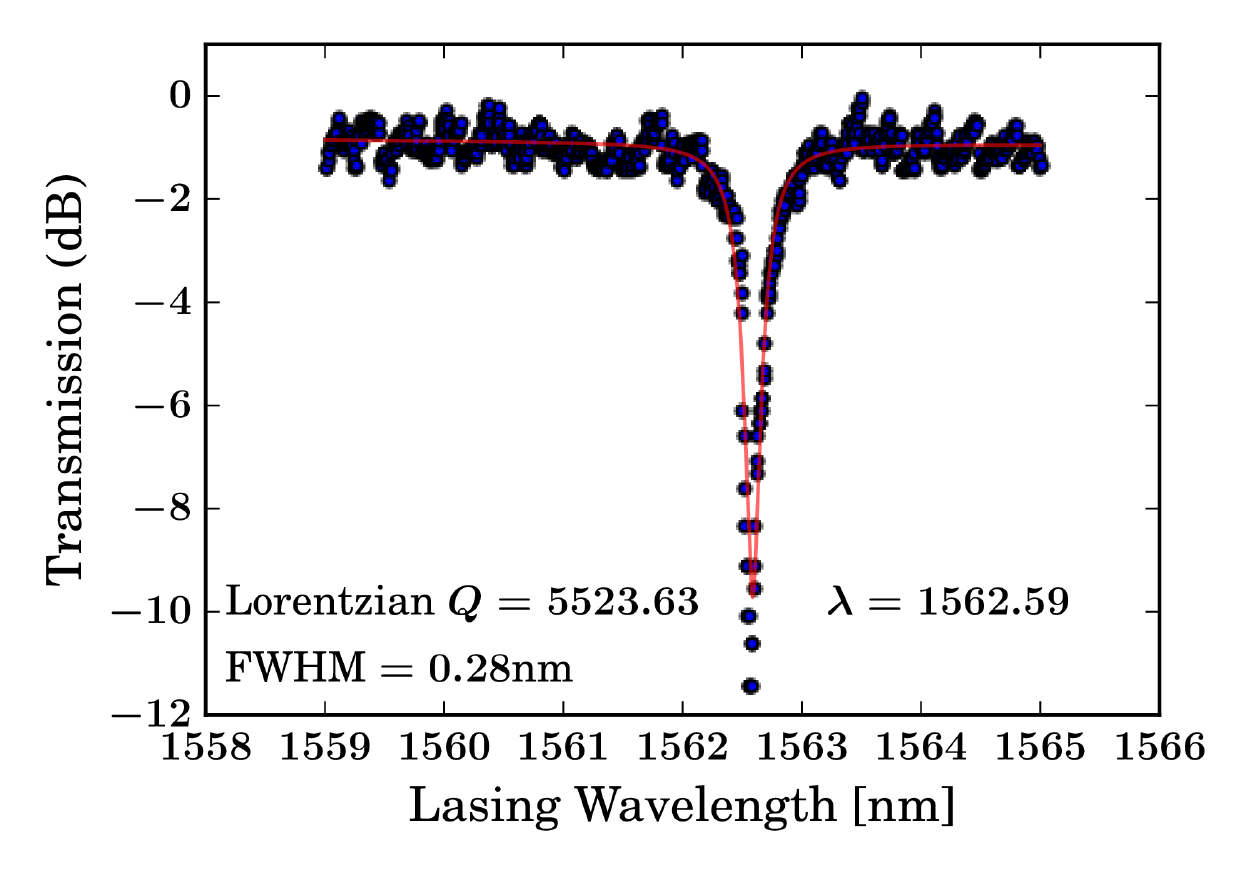}
\caption{Example of a measured notch for a 25$\mu$m radius Si$_{3}$N$_{4}$ ring device.  The parameters from a Lorentzian fit to the data are also shown.}
\label{fig:notchfits}
\end{figure}

\begin{table}[ht]
\begin{center}
\caption{Ring Resonator designs tested successfully so far.}
\label{tab:tested}
\begin{tabular}{|c|c|c|}
\hline
Basic Design & Material & Number of Tested Devices \\
\hline\hline
Single 25um ring & SiN & 7 \\
Single 25um ring & Si & 1 \\
Single 10um ring & SiN & 4 \\
Single 5um ring & SiN & 3 \\
Racetracks & SiN & 3 \\
Double Rings in Series & SiN & 5 \\
\hline
Total & & 23 \\
\hline
\end{tabular}
\end{center}
\end{table}

The test procedure at Macquarie University uses a JDS Uniphase Swept Wavelength System (SWS) to measure transmission over the wavelength range 1.52 -- 1.57~$\mu$m.  This is injected into the waveguides via lensed fibres with a 2.5~$\mu$m spot-size, which are aligned to the waveguides with Melles-Griot piezo-controlled six-axis stages.  The light is fed from the waveguide back to the SWS using an identical procedure.

The test procedure at Argonne uses a Thorlabs TLK-L1550R tunable laser to measure transmission over the wavelength range 1.48 -- 1.62~$\mu$m.  The same type of lensed fibres used at Macquarie University inject light into the waveguides, after alignment with five-axis stages.  The light from the output waveguide is detected with an IR photodiode and recorded with custom electronics. 

The results of the lab tests are discussed below, in terms of their relevance to the OH suppression requirements laid out in section~\ref{sec:ohreq}.

\subsubsection{Notch depth results}

The deepest notch measured so far from the tested devices is shown in Figure~\ref{fig:notchfits}, with a relative transmission of $\approx -9.5$~dB , corresponding to a suppression factor of  8.9.

Comparing to Figure~\ref{fig:nnotches}, for $>60$ notches across the J and H bands, an average suppression of $-8.7$~dB would improve the signal to noise by a factor of  $\approx 2.5$ if the notches are 100 - 200~pm wide.  Note that so far no tuning of the coupling coefficients has been made.  Future tuning to ensure $t \approx \alpha$ should ensure deeper notches.

\subsubsection{Free spectral range results}
\label{sec:fsrresults}

\begin{figure}[ht]
\centering \includegraphics[scale=0.8]{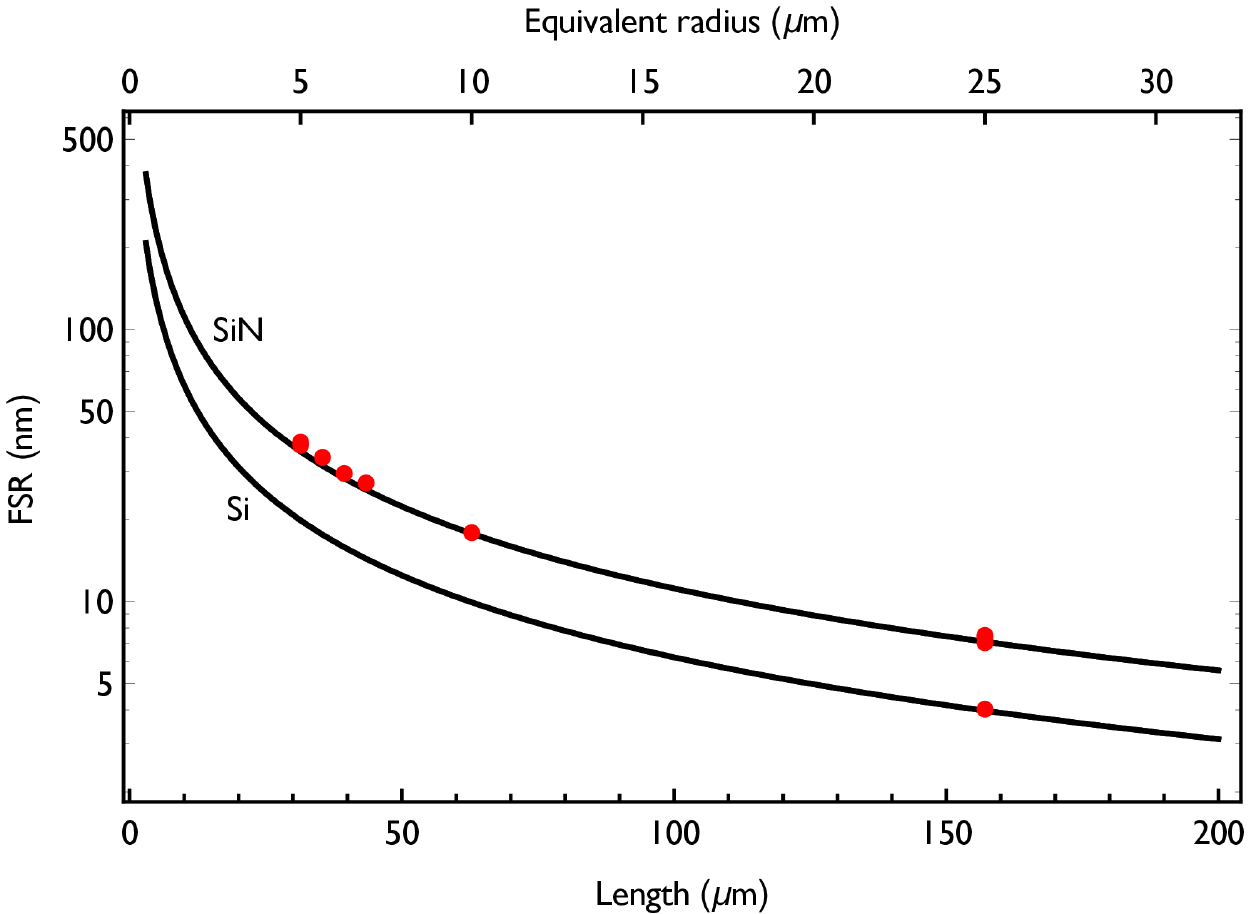}
\caption{The measured FSRs (red points) of several devices compared to the theoretical curves for Si$_{3}$N$_{4}$ and Si, with $n_{\rm g} = 2.15$ and $n_{\rm g} = 3.85$ respectively.}
\label{fig:fsrdata}
\end{figure}

The free spectral ranges (in nm) of all measured devices are shown in Figure~\ref{fig:fsrdata}, as a function of the total length of the resonator (bottom x-axis), and the equivalent ring radius (top x-axis).  The data (red points) include rings and racetrack geometries,  as well as both silicon nitride and silicon waveguides.  The data are in good agreement with the theoretical curves.  The FSR was found not to depend strongly on the polarisation, and we measure the FSR with a single polarisation state.  
The group index was determined from fits to the FSR of one Si or  Si$_{3}$N$_{4}$ device, and used to produce the theoretical curves at all resonator lengths in Figure~\ref{fig:fsrdata}.

\subsubsection{Notch tuning and accuracy results}
\label{sec:doubleringresults}

The electron beam lithography system we are using to fabricate resonators has a precision of better than 10~nm,  which by itself is not precise enough to meet our absolute notch wavelength specification of 0.1nm, discussed in section~\ref{notchspec}. However, the more important test of the system is to predict the \textit{differences} in wavelengths between multiple rings,  and this has never been tested before.  If these differences cannot be reliably predicted,  then a system of individually tuned rings with temperature or applied voltage will be required.     

\begin{figure}[ht]
\centering 
\includegraphics[scale=0.3]{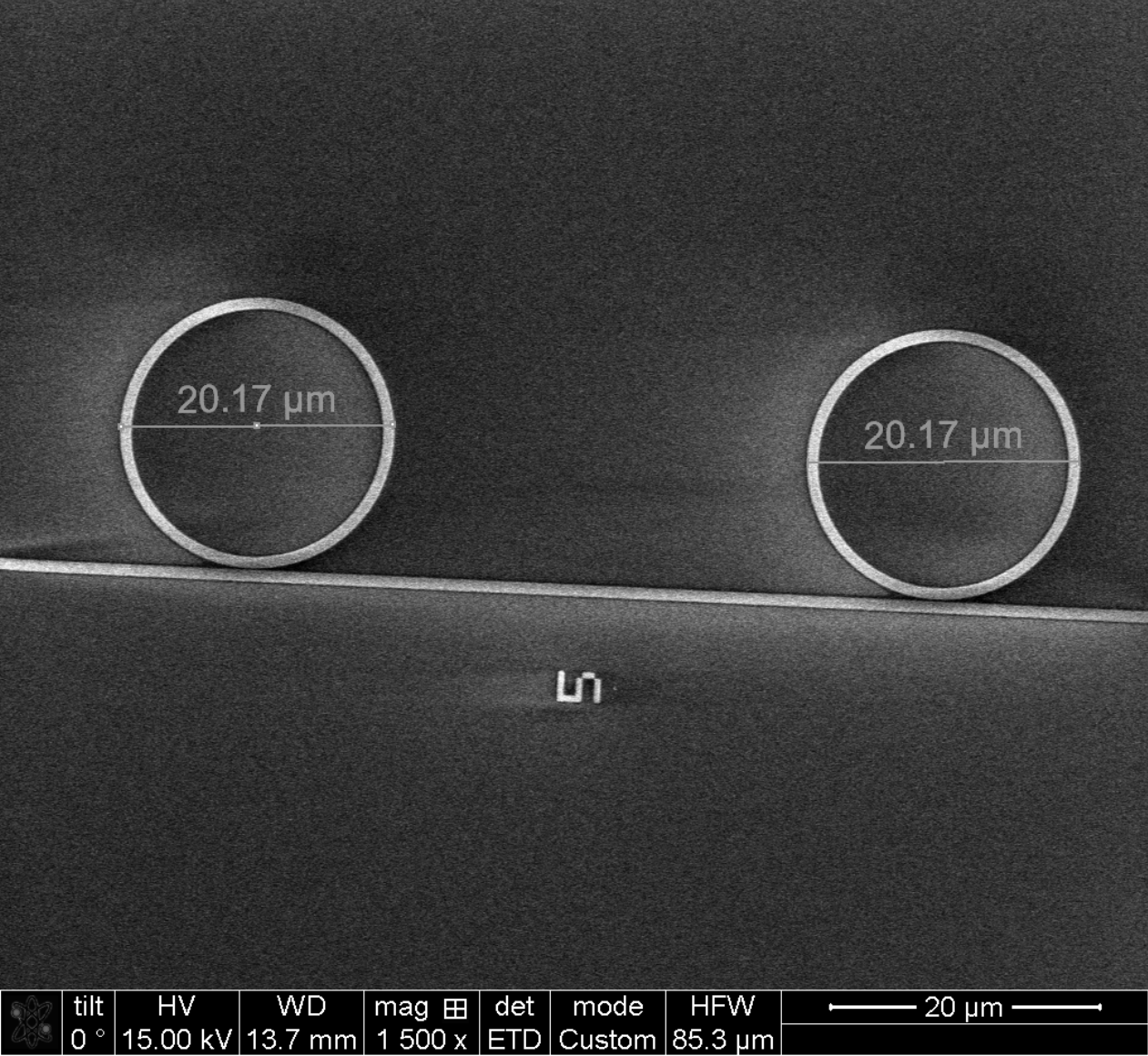}
\includegraphics[scale=0.4]{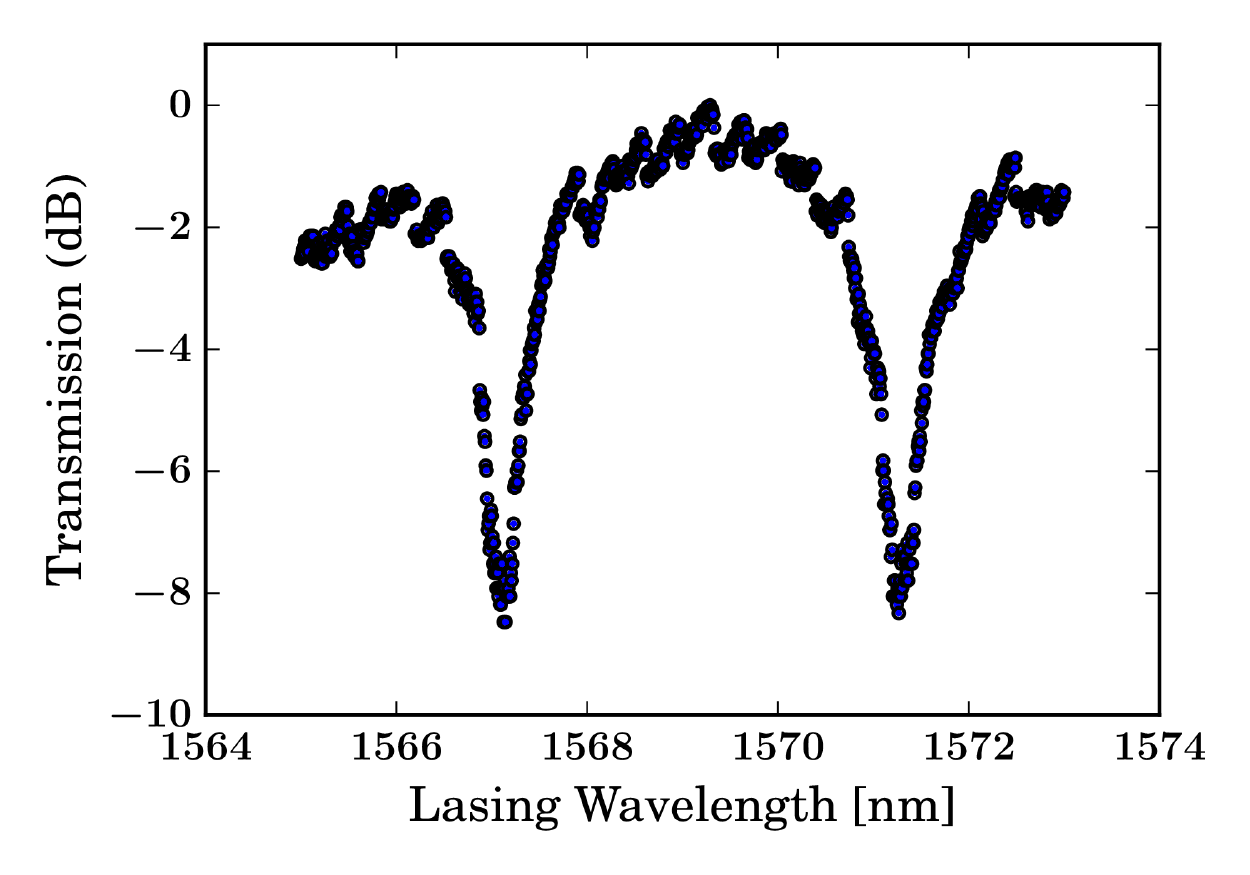}
\caption{(Left panel.) SEM photograph of two rings in series, with design radii 10$\mu$m and 10.032$\mu$m (too small to discern in the SEM).  (Right panel.) The spectrum of this two-ring system, producing two notches with a difference in wavelength consistent with 3D simulations.} 
\label{fig:notchdata}
\end{figure}

As an initial test,  we have fabricated five devices with two rings in series on the same waveguide.  One of these devices is shown in the left panel of Figure~\ref{fig:notchdata}, with one ring radii designed to be 10~$\mu$m, the other ring radii designed to be 10.032~$\mu$m.  

We used 3D FDTD simulations with Lumerical to  predict the notch wavelength differences.  These were 3.97~nm to 4.02~nm, depending on polarization.  The right panel of Figure~\ref{fig:notchdata} shows the response of this device,  and fits to the notches give a difference of 3.96~nm, very close to the 3D simulated value.   A replica of this ring system on a different wafer gave a difference of 4.14~nm,  also very close to the simulated value.   The other three double rings had much smaller radii differences, but produced multiple notches consistent with expectations. We plan to produce dozens of multi-ring systems to give sufficient statistics to understand the RMS spread of the fabrication uncertainty.

\section{Discussion}
\label{sec:discuss}

We have examined the application of ring resonators to astronomical instrumentation, with a focus on their use as notch filters for OH suppression.  Ring resonators are an attractive solution to the problem of OH suppression, because their size and method of fabrication is ideally suited to incorporating hundreds of rings in a single monolithic device\cite{ell12c}.  Thus they are very scalable to the demands of a scientific instrument, and can be incorporated into an instrument in a modular fashion.  

This modularity will become increasingly important in the era of extremely large telescopes.  The beam size of a telescope scales with its diameter, and so as telescopes get larger so must astronomical instruments.  However, this scaling can be broken by working at the diffraction limit\cite{bland10}, when the beam size is independent of telescope aperture.  Thus by splitting the multimode beam of a seeing limited telescope into many single mode fibres via a photonic lantern, the problem changes from one of increasing the size of the instruments to one of replicating many miniature modular devices.  The lithographic manufacture of ring resonators makes them ideally suited for this task.

We reviewed the astronomical requirements for OH suppression in section~\ref{sec:ohsupp}, and thereafter discussed the consequent requirements for ring resonators in section~\ref{sec:ohreq}.  From a theoretical perspective ring resonators are very well suited for OH suppression provided that certain conditions can be met.  In particular they must have low loss, and optimally matched coupling coefficients, i.e.\ $\alpha=t > 0.9$.  If this can be achieved, then the requirements of deep notch depth ($< - 30$~dB), narrow notch width ($<200$~pm) and high interline throughput ($>0.8$) can all be met.  

Another requirement is to achieve a large FSR ($> 30$~nm), such that within the wavelength range of interest there is only one notch from each ring.  Larger wavelength ranges can be built-up from several such $\sim 30$~nm windows.  A large FSR requires a small radius ring.  Therefore care must be taken to avoid large bend losses.  This leads us to waveguides with large refractive indices, such as Si or Si$_{3}$N$_{4}$; although note that a large index reduces the FSR, so a balance must be found.

Starting from these theoretical requirements we have fabricated several Si and Si$_{3}$N$_{4}$ waveguides in order to test their performance and to determine the design and fabrication specifications.  Although this work is yet at an early stage we have  several promising preliminary results.

First, we have measured FSRs for several devices (Fig.~\ref{fig:fsrdata}), and find results fully consistent with the theoretical performance of our devices.  Moreover, we have already fabricated devices with FSR $> 30$~nm, meeting the goal developed in section~\ref{sec:ohreq} for simple ring resonator devices.

Second, we have fabricated devices with double rings, with different resonant wavelengths (Fig.~\ref{fig:notchdata}).  Again, the measured resonant wavelengths are fully consistent with the theoretical predictions.

Third, we have measured the quality, throughput ($\alpha$), and self-coupling coefficients of several rings.  We find that the $\alpha$/$t$ parameters are $>0.9$, which is high enough to ensure good interline throughput over the 30~nm wavelength range.  However, these $\alpha$/$t$ values are not yet high enough to ensure adequate $Q$, and indeed the measured $Q$s of $\approx 4000$ do not yet have high enough resolution for OH suppression.  However, it is important to note that for these devices we made no attempt to  tune the coupling.   It is entirely reasonable to expect that an investigation of the coupling parameters, e.g.\ tuning the gap size, coupling length or race-track/ circular resonators, as well as the waveguide dimensions, will yield devices with much higher $Q$, such as those seen in the literature (e.g.\ $Q\approx 45000$ for a 30~$\mu$m Si resonator\cite{bae04} or $Q\approx 9000$ for a 1.5~$\mu$m Si resonator\cite{xu08}).

The next stages of development for ring resonator based OH suppression requires progress in three main areas.  First, we need to optimise the behaviour of the rings themselves.  That is, we need to identify and tune the design parameters to increase $\alpha$/$t$ and $Q$.  Devices to test these parameters are already being made, and these tests are ongoing. 

Second, we need to demonstrate the tuning of multi-ring devices to the specific OH wavelengths.  Since we can already predict the accurate spacings of double ring devices, the printing of multi-ring devices with correct spacings should not be a fundamental barrier.  However, there is a second part of this step, which requires the absolute tuning of the ring frequencies to the OH lines, and thereafter maintaining this tuning.  This should be accomplished using temperature control of all the rings simultaneously.  We have not yet tested the temperature control of the rings, but the requirements on the stability are not expected to be severe (\S~\ref{sec:stability}).

Third, we need to demonstrate efficient fibre-chip coupling.  This is currently the biggest challenge for incorporating ring resonators into astronomical instruments.
The fundamental mismatch between the MFD of the waveguides and that of a SMF adds an extra level of difficulty in accepting light from the focus of a telescope, since not only must there be a multimode to single mode conversion, but also a fibre to chip conversion.   We are currently investigating the use of inverted tapers butt-coupled to lensed fibres, and have previously fabricated V-groove aligned cantilevers (\S~\ref{sec:injection}).  We note that this same problem presents a  challenge to the use of of silicon photonics in general\cite{tho16}, and hence it has received considerable attention, since solving it would enable the CMOS manufacture of many photonic components.  However, the challenges for commercial packaging are somewhat different, with high-volume and low-cost automated packaging a priority.  For astronomical instruments we do not need high volume, since instruments are built on a case-by-base basis, and for the same reason we do not necessarily require automated fibre-chip packaging.  However, we do require very high efficiency, since the astronomical targets of interest are by their nature intrinsically faint, and ring resonators will be one part of a long chain of optical components from the telescope to the detector.

In summary, the overall prospects for ring resonator based astronomical instruments are good.  The internal properties of the devices present no fundamental barriers to their use, and it should be possible to meet all the requirements for OH suppression, with the caveats that the large wavelength range will need to be split over several sub-windows, and there will need to be multiple copies of each window for every spatial and polarisation mode.  However, the lithographic fabrication of ring resonators makes such replication trivial.  
There is also reason to be optimistic about the integration of ring resonators to astronomical instruments despite the current challenges in coupling to fibres.  Coupling techniques are an active area of research, and moreover we can afford to spend the time 
fine tuning the most promising techniques, since we do not require mass production.

Of course, all these issues will take considerable development before  a ring resonator based instrument is realised.  However, this development is well worth the investment.  The challenge of building affordable instruments in the current era with extremely large telescopes is not trivial.  Astrophotonic instruments based on optical devices embedded within single mode waveguides offer a new solution to the process of building instruments for large telescopes.  Already there are astrophotonic components of traditional instruments, providing new and better functionality, e.g.\ interferometric beam combination\cite{ker97}, pupil re-mapping aperture masking\cite{jov12}, FBG OH suppression\cite{ell12a,tri13a,bland11b}.  However, the true potential of astrophotonics lies in fully photonic instruments.  Once the light from the telescope is fed into single mode fibres, all subsequent processing of light can take place on photonic chips.  For example, following ring resonator based OH suppression spectroscopy could be carried out on the same chip using array waveguide gratings\cite{cve09,cve12a,cve12b,bland10}, since the same fabrication methods can be used for both.  This would lead to truly modular and miniature instruments.

\section*{Funding}

Argonne National Laboratory's work was supported under U.S.
Department of Energy, Office of Science, contract DE-AC02-06CH11357.  Use of the Center for Nanoscale Materials, an Office of Science user facility, was also supported by the U. S. Department of Energy, Office of Science, Office of Basic Energy Sciences, under the same contract listed above.  Work by G. Wei was supported by the U.S. Department of Energy (DOE), Office of Science, Basic Energy Sciences under Award DE-SC0012130.  N.P. Stern acknowledges support as an Alfred P. Sloan Research Fellow.

This work was performed in-part at the OptoFab node of the Australian National Fabrication Facility, utilising NCRIS and NSW state government funding. 

\section*{Acknowledgments}

Many thanks to Martin Ams, Ben Johnston, Alex Stokes, and Graham Smith for their invaluable help in the laboratory testing of our devices.  We are also very grateful for the help of our AAO vacation students:
R.\ Nash, J.\ St.\ Antoine, J.\ Lorenzo Redondo, J.\ Kepple, and A.\ Crouzier.  Many thanks to Jon Lawrence and Sergio Le\'{o}n-Saval for useful discussions.

We wish to thank the staff at the Argonne Center for Nanoscale Materials,  whose contributions made the ring resonator fabrication possible. This includes co-author Leo Ocola, Dave Czaplewski, Ralu Divan, Suzanne Miller,  and Valentina Kutepova.   Argonne students have been crucial for our device testing program:  Danny Davies, James DerKacy, Ariel Matalon, Alexis Miranda, Kasia Pomian, and Joe Pastore. We wish to thank Tom Kasprzyk for his technical help with the Argonne test-stand.

\appendix

\section{Other uses of ring resonators for astronomy}
\label{sec:other}

In this paper we have concentrated on the use of ring resonators for OH suppression, since that is where our main efforts have been focussed to date.  However, ring resonators are potentially useful for other astronomical applications which we shall briefly review here, without developing the detailed requirements.

\subsection{Wavelength calibration}
\label{sec:freqcomb}

If the input port of an add-drop resonator is fed with a continuous-wave continuum source the ring will naturally generate a frequency comb at the drop port.   This frequency comb can provide a potentially useful wavelength calibration source, especially for fibre-fed spectrographs or photonic instruments, which may be easily integrated with a waveguide source.

The major focus on using laser frequency combs in astronomy has been to generate extremely high accuracy calibration sources  for precision radial velocity measurements in planet search experiments\cite{ste08,mur12,lee12}.  Ring resonators can be used to generate similar combs using the non-linear effect of four-wave mixing; if a ring with resonant wavelength $\lambda_{0}$ is fed with a powerful laser of wavelength $\lambda_{1}$, non-linear effects in the material of the ring will generate a third wavelength at a wavelength of $\lambda_{2} = 2 \lambda_{1} - \lambda_{0}$.  Provided the source  is powerful enough, the newly generated line may also mix with the existing lines to create a further lines at  $\lambda_{3} = 2 \lambda_{2} - \lambda{1}$, $\lambda_{4} = 2 \lambda_{2} - \lambda_{0}$, etc.\ and in this way a frequency comb is generated with a wavelength spacing of $\Delta \lambda = \lambda_{1} - \lambda_{0}$.

The advantages of using four-wave mixing to generate frequency combs are that all the power from the comb is in the frequency peaks, with no power in between, the comb is regularly spaced in wavelength, and the light from  all frequencies are in phase.   The disadvantage is that in order to exploit the non-linear effects high power lasers and materials with highly non-linear properties must be used.  This approach is being pursued in photonics research\cite{kip11} and in astronomical applications\cite{cha12}.

Here we wish to concentrate on the simpler method of frequency comb generation using passive ring resonators,
with the more modest goals of moderate resolution wavelength calibration, such as is typically achieved with arc lamps.  Ring resonators have the advantage over arc lamps that the FSR may be designed or tuned to give as many frequencies as desired, and that  similar intensities may be achieved over all frequencies.   Moreover, the frequency spacing of the combs generated by typical rings of radius tens or hundreds of microns, naturally provide a suitable spacing of $\approx 10$ times the resolution element for resolving powers of 100 to 100,000; this is not the case for laser-frequency combs which must be subsequently filtered to provide a suitable comb spacing. 

\subsubsection{Requirements}

The FSR is required to provide a sufficient 
number of lines to allow accurate calibration, whilst maintaining adequate separate between the lines such that there is no significant overlap.  For example, for a NIR spectrograph with a resolving power of $R=3000$, the resolution of the spectrograph is $\approx 0.5$~nm, thus a FSR of $\approx 5$~nm would provide $>50$ well separated lines.  The exact requirements will depend on the spectrograph, but  with a judicious choice of radius and materials for the core and cladding FSRs from 0.3 to 100 nm are feasible.

The emission from the frequency comb between the lines should be as close to zero as possible.  (Note for the case of four-wave-mixing, this condition will always be met, hence its appeal.)   This is best achieved by ensuring that the cross-coupling coefficient is small (i.e.\ $t$ is high).

The line width of the frequency comb should be narrow enough such that the measured line width and shape is entirely determined by the spectrograph.  In this way the PSF of the spectrograph can be calibrated across all wavelengths.

\begin{figure}
\centering \includegraphics[scale=0.65]{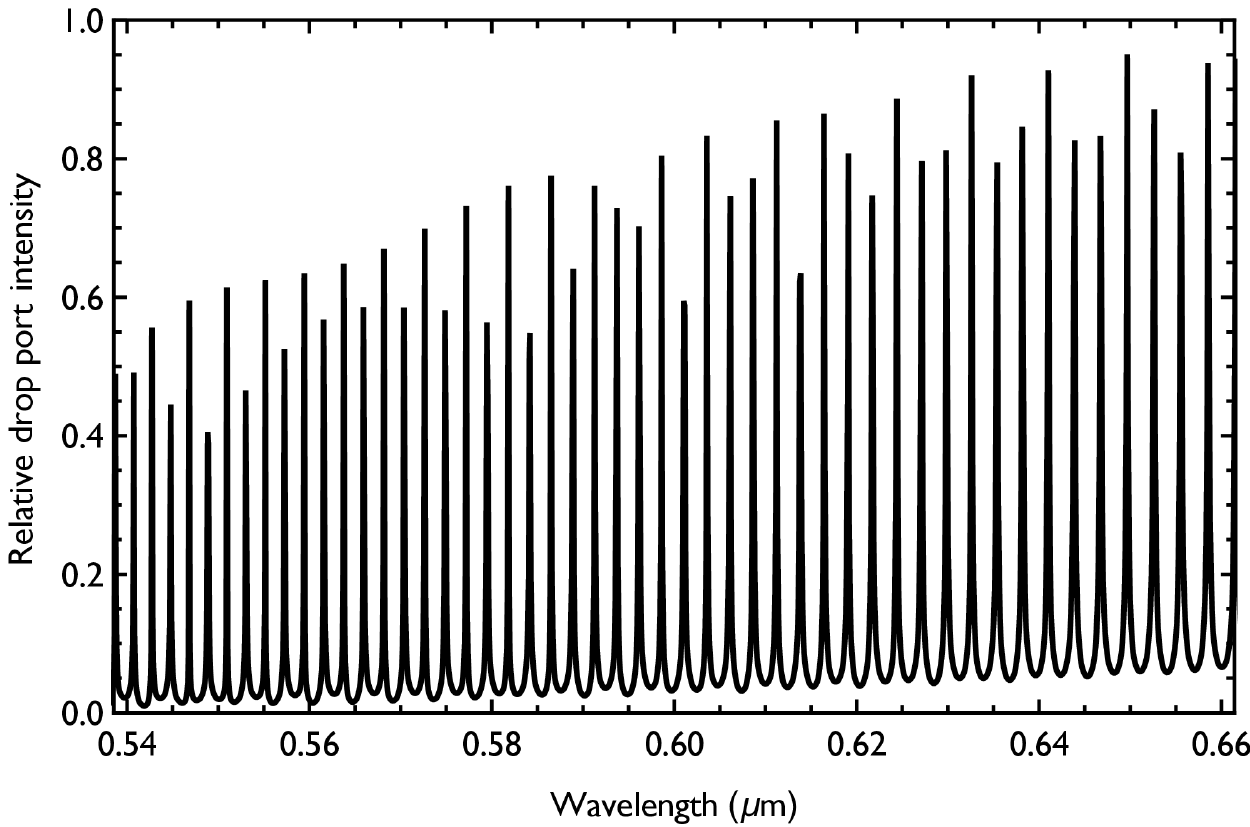}
\caption{The relative intensity at the drop port for a 2D FDTD simulation of the frequency comb for the R=10000 visible spectrograph with $FSR\approx 2.5$~nm.   This would provide a wavelength calibration accurate to 0.4 pixels or 0.1\AA.}
\label{fig:freqsim}
\end{figure}

The quality of a ring resonator (and also the bandwidth) is a function of the group index of the waveguide, the ring circumference, the coupling coefficients, the losses in the ring and the wavelength.  For high quality $n_{\rm g}$ and $L$ should be large, and $\alpha$ and $t$ should be close to unity.  However, the ring circumference and material are also determined by the requirements on the FSR, and the coupling coefficient is determined by the requirement on the interpeak emission.  Likewise, the ring losses are determined by the ring circumference and material.  That is, all of the parameters which determine the quality of the ring, are also constrained by other requirements.  Happily, all these constraints are usually complementary for frequency combs, which do not generally require large FSR., see e.g.\ Figure~\ref{fig:freqsim}, showing a 2D  {\sc RSOFT FullWAVE} FDTD simulation of 
a Si$_{3}$N$_{4}$ ring resonator of radius $\approx$11.1~$\mu$m to provide a $FSR\approx 2.5$~nm comb for a $R=10000$ spectrograph at $\lambda = 600$~nm.

The stability of the comb wavelengths is important to achieve accurate calibration.  In section~\ref{sec:stability} we have shown that the requirements for achieving adequate stability are not onerous due to the weak temperature dependence of typical devices, and their small size.  The same is true here, and we note that the results of Lee et al.\cite{lee12}, which demonstrated 1~pm accuracy over a 24 hr period were for a device designed as a frequency comb.

\subsection{Doppler planet search}
\label{sec:dopp}

When a planet orbits a star, both star and planet revolve around the centre of mass of the system.  Thus the position of the star `wobbles' by an amount that depends on the mass and orbital radius of the planet.  This `wobble' is observable as a Doppler shift in the wavelength of the stellar spectrum.  Such Doppler planet searches are a well studied and extremely fruitful way of finding extrasolar planets\cite{may14}.

Doppler planet searches require high resolution spectroscopy in order to see the small shifts ($< 10$~m~s$^{-1}$) associated with planets.  High resolution spectroscopy is restricted to relatively bright targets: the light is highly dispersed, therefore the number of photons/s/pixel  is low, and correspondingly the detector pixel noise is more significant.  Lower mass systems require higher precision, and are therefore even more limited in sensitivity.  Furthermore much of the recorded signal is redundant, since only those parts of the spectrum with measurable features contribute to the velocity measurement.

\begin{figure}
\centering \includegraphics[scale=0.4]{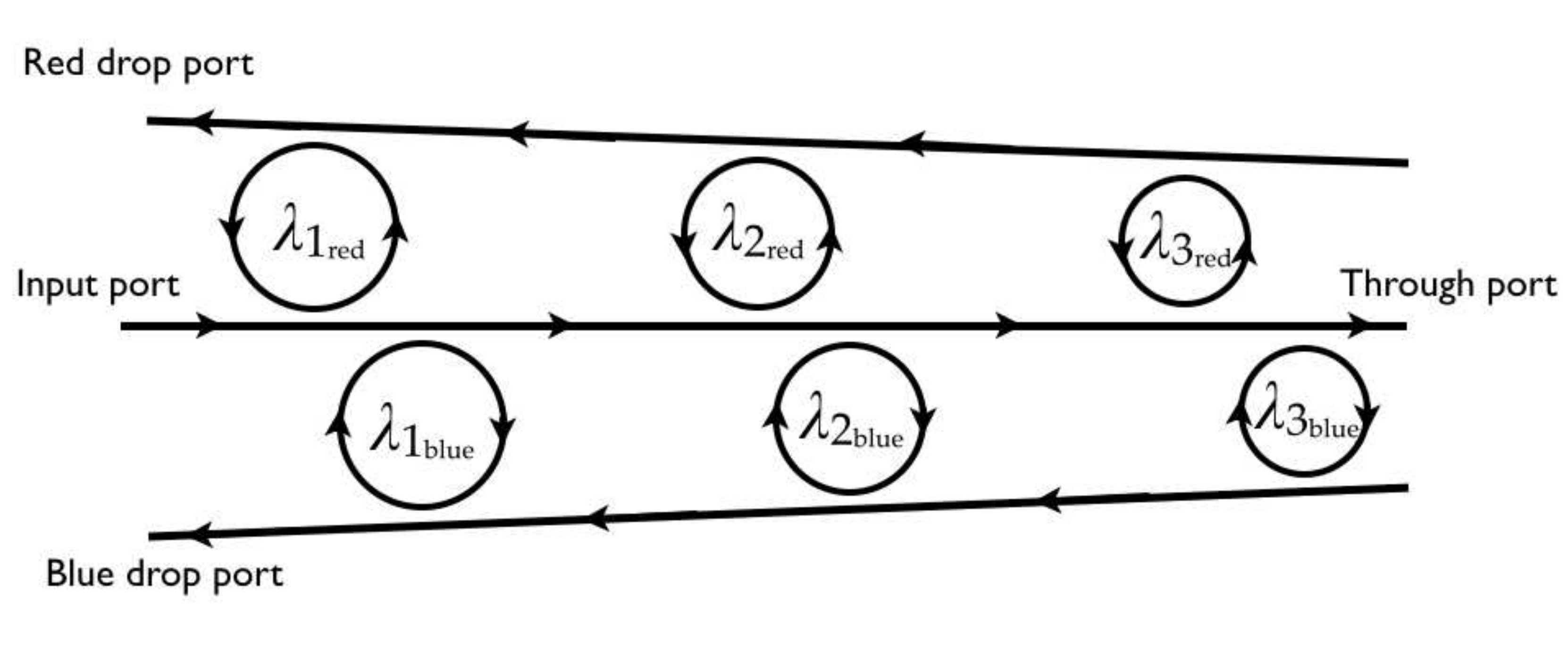}
\caption{Schematic diagram of a conceptual Doppler planet search ring resonator circuit based on three spectral features.}
\label{fig:dopp}
\end{figure}

Ring resonators offer an interesting potential alternative to high resolution spectroscopy for Doppler planet searches.  Consider Figure~\ref{fig:dopp}.  The input waveguide has pairs of rings on either side of it, each feeding a drop port.  Each pair of rings is tuned in wavelength to a particular spectral feature of the stellar spectrum, but one of the pair is tuned slightly blue and the other slightly red.  For example, since most planet searches take place around cool long-lived stars, the spectra will contain many absorption lines.  Each pair of rings may be tuned to wavelengths just either side of an absorption line.
Thus as the stellar spectrum is Doppler shifted to the red and the blue the signal from the absorption line is alternately shifted into one, then the other ring.  Each pair of rings will do the same for a different absorption line.

The key point is that the dropped signals from each of the  rings are added together \emph{photonically}, prior to any electronic detection.  There is no extra electronic noise associated with adding the signals together, as there would be if they were recorded in a spectrum.  The noise of the added signals is simply the Poisson noise of each signal added in quadrature.  Furthermore, the added signals can then be detected by three single pixel detectors, one for each drop port, and another at the through port for photometric calibration.  As the planet orbits the star, the signal would decrease periodically  and alternately in each of the detectors as the absorption features are shifted from one drop port to the other.

Therefore, ring resonators allow many Doppler shifted spectral features to be added together, selecting only the high signal parts of the spectrum, with no unwanted featureless parts, and with minimal electronic detector noise.

\subsubsection{Requirements}

Here we give only a brief consideration of the requirements.  The idea presented here is only conceptual, and a full appraisal of the benefits and requirements will require detailed modelling of the instrument and Doppler shifted spectra, and will be developed in a future paper.
In general the requirements are similar to those of OH suppression, since many features must be simultaneous filtered, without unwanted interference from secondary resonances.  One significant difference from OH suppression is that Doppler searches are best carried out at visible wavelengths.

The FSR must be large enough so that each ring only filters one line  in the wavelength of interest, otherwise the dropped signal will be contaminated with unwanted parts of the spectrum.  The FSR therefore depends on how many lines are required to be measured, and their spacing.  For comparison, HARPS has a bandpass from $0.38$ -- $0.69$~$\mu$m, and thus it may be necessary to have separate sub-windows as for OH suppression.
 
The accuracy of the Doppler shifts measured depends on the stability and calibration of the notch transmission profiles.  As for high resolution spectroscopy, the accuracy can be much higher than the $\Delta \lambda$ resolution of the instrument (cf.\ UCLES on the AAT has $R\approx 100,000$, but achieves accuracies of $\approx 3$~m~s$^{-1}$, that is $\approx 1000$ times better than the $\Delta \lambda$ resolution).  The same will be true for resonators, and the exact precision will require detailed modelling, but it is clear that high $Q$ factors will be necessary.

\end{document}